# Revealing Callisto's carbon-rich surface and $CO_2$ atmosphere with JWST

*Richard J. Cartwright[1], Geronimo L. Villanueva[2], Bryan J. Holler[3], Maria Camarca[4], Sara Faggi[3], Marc Neveu[5,3], Lorenz Roth[6], Ujjwal Raut[7], Christopher R. Glein[7], Julie C. Castillo-Rogez[8,4], Michael J. Malaska[8,4], Dominique Bockelée-Morvan[9], Tom A. Nordheim[1], Kevin P. Hand[8,4], Giovanni Strazzulla[10], Yvonne J. Pendleton[11], Katherine de Kleer[4], Chloe B. Beddingfield,[1], Imke de Pater[12], Dale P. Cruikshank[10], Silvia Protopapa[7]*

*[1]Johns Hopkins University Applied Physics Laboratory, [2]NASA Goddard Space Flight Center, [3]Space Telescope Science Institute, [4]California Institute of Technology, [5]University of Maryland, [6]KTH Royal Institute of Technology, [7]Southwest Research Institute, [8]Jet Propulsion Laboratory, [9]Observatoire de Paris, CNRS, [10]Istituto Nazionale di Astrofisica, [11]University of Central Florida, [12]University of California Berkeley*

**Abstract**

We analyzed spectral cubes of Callisto's leading and trailing hemispheres, collected with the NIRSpec Integrated Field Unit (G395H) on the James Webb Space Telescope. These spatially resolved data show strong 4.25-µm absorption bands resulting from solid-state $^{12}CO_2$, with the strongest spectral features at low latitudes near the center of its trailing hemisphere, consistent with radiolytic production spurred by magnetospheric plasma interacting with native $H_2O$ mixed with carbonaceous compounds. We detected $CO_2$ rovibrational emission lines between 4.2 and 4.3 µm over both hemispheres, confirming the global presence of $CO_2$ gas in Callisto's tenuous atmosphere. These results represent the first detection of $CO_2$ gas over Callisto's trailing side. The distribution of $CO_2$ gas is offset from the subsolar region on either hemisphere, suggesting that sputtering, radiolysis, and geologic processes help sustain Callisto's atmosphere. We detected a 4.38-µm absorption band that likely results from solid-state $^{13}CO_2$. A prominent 4.57-µm absorption band that might result from CN-bearing organics is present and significantly stronger on Callisto's leading hemisphere, unlike $^{12}CO_2$, suggesting these two spectral features are spatially anti-associated. The distribution of the 4.57-µm band is more consistent with a native origin and/or accumulation of dust from Jupiter's irregular satellites. Other, more subtle absorption features could result from CH-bearing organics, CO, carbonyl sulfide (OCS), and Na-bearing minerals. These results highlight the need for preparatory laboratory work and improved surface-atmosphere interaction models to better understand carbon chemistry on the icy Galilean moons before the arrival of NASA's Europa Clipper and ESA's JUICE spacecraft.

## 1. Introduction

The Galilean moon Callisto has one of the most ancient surfaces in the Solar System (>4 Ga, *e.g.*, Zahnle et al., 2003). As first seen by Voyager during its flyby of the Jovian system in 1977, the surface of Callisto is heavily cratered with minimal evidence for endogenic modification, unlike the other Galilean moons that each display ubiquitous evidence for recent resurfacing. The arrival of the Galileo orbiter in 1996 revealed the surfaces of the Galilean moons in stunning detail (*e.g.*, Showman and Malhotra, 1999). Observed during multiple close flybys, Callisto's surface geology is dominated by heavily degraded craters, large mass wasting deposits, and fields of bright, icy pinnacles protruding from a smooth blanket of dark material that is pervasive across Callisto's surface (Moore et al., 2004).

Near-infrared, ground-based observations determined that Callisto's surface is primarily composed of $H_2O$ ice mixed with a dark component that could include organics, phyllosilicates, and other hydrated minerals (*e.g.*, Pollack et al., 1978; Clark et al., 1980; Calvin and Clark, 1991), such as ammonium ($NH_4$) bearing compounds (Calvin and Clark, 1993). Reflectance spectra collected by the Near Infrared Mapping Spectrometer (NIMS) on Galileo confirmed the presence of $H_2O$ ice mixed with dark material, and also provided the first detection of sulfur-bearing species and large amounts of solid-state $CO_2$ (Carlson et al., 1996). Furthermore, NIMS detected subtle absorption features between 3.3 and 3.5 μm that may result from C-H stretching modes of short chain aliphatic organics, as well as broad bands centered near 3.88, 4.02, and 4.57 μm, likely resulting from carbon and/or sulfur-bearing species (McCord et al., 1997, 1998a). The possible presence of these components suggests that dark material on Callisto includes amorphous carbon and complex organic residues perhaps similar to 'tholins' generated in the laboratory (*e.g.*, Cruikshank et al., 1991; Khare et al., 1993). In an off-limb scan above the leading hemisphere, NIMS also detected gaseous emissions from a tenuous $CO_2$ atmosphere, 5 to 40 km above Callisto's surface (Carlson, 1999), further highlighting Callisto's C-rich environment.

The large amount of $CO_2$ on Callisto's surface, the presence of $CO_2$ gas, and a mean subsolar temperature of 165 K (Spencer, 1987a), at which $CO_2$ ice on Callisto would be thermodynamically unstable (*e.g.*, Brown and Ziegler, 1979; Fray and Schmitt, 2009, and references therein), have been implicated in crater degradation and icy pinnacle formation processes on Callisto. In this scenario, $H_2O$ ice and crustal $CO_2$ ice sublimate, spurring the disaggregation of crater rims and triggering large mass wasting events (Moore et al., 1999, 2004; Howard and Moore, 2008; White et al., 2016). $CO_2$ molecules sublime and are then transported to ice-rich, reflective terrains where they might condense in cold traps such as $H_2O$ ice rich remnant crater rim segments and other high standing terrains. Callisto's ubiquitous dark material could therefore result from widespread, sublimation-driven erosion of crustal sources of $CO_2$, building up a lag deposit rich in carbonaceous material mixed with hydrated minerals.

Exogenic processes are also likely contributing to Callisto's surface inventory of $CO_2$ and other carbon oxides. Charged particle interactions with C-rich, icy deposits on Callisto's surface drives radiolytic chemistry, likely resulting in a carbon cycle, including production of $CO_2$ molecules. The 'bullseye' shaped distribution of solid-state $CO_2$ detected by NIMS, with band depths peaking near Callisto's trailing side apex, is consistent with $CO_2$ generated by radiolytic interactions between fast rotating plasma in the Jovian magnetosphere and C-rich material and $H_2O$ ice in Callisto's regolith (*e.g.*, Hibbitts et al., 2000). A wide variety of laboratory experiments demonstrate that $CO_2$ is efficiently produced at the interface between solid elemental carbon (amorphous carbon, residues from energetic processing of C-bearing ices, bitumens, *etc*) and $H_2O$ ice (*e.g.*, Spinks and Wood, 1990; Mennella et al. 2004; Gomis and Strazzulla 2005; Strazzulla and Moroz 2005; Raut et al., 2012). In contrast, solid-state $CO_2$ on Callisto's leading hemisphere appears to be spatially associated with craters and their ejecta, consistent with a crustal source of $CO_2$ (Hibbitts et al., 2002) and/or cold trapping of radiolytically-produced $CO_2$ on relatively bright crater floors, rims, and ejecta blankets. Supporting a native origin for $CO_2$, satellite formation models indicate that Callisto accreted large amounts of this molecule as it formed in the Jovian subnebula (*e.g.*, Mousis and Alibert, 2006; Melwani Daswani et al., 2021). Whether radiolytic or crustal sources dominate Callisto's observed surface inventory of $CO_2$ is uncertain.

The solid-state $CO_2$ feature is centered near 4.258 μm in NIMS data (Carlson et al., 1996), notably offset from the asymmetric stretch fundamental ($\nu_3$) of crystalline $CO_2$ ice measured in the laboratory (~4.27 μm) (*e.g.*, Quirico and Schmidt, 1997; Hansen, 1997). This 4.27-μm band is exhibited by $CO_2$ ice produced by energetic processing of C and O-bearing frozen gases, as well as after energetic processing of $H_2O$ ice deposited on top of solid carbonaceous materials at low temperatures (<100 K) (*e.g.*, Raut et al., 2012; Jones et al. 2014). $CO_2$ ice is not stable at Callisto's peak surface temperatures, and instead $CO_2$ molecules are likely bound to more refractory components such as dark material possibly including salts (Villanueva et al., 2023a) or $H_2O$ ice (McCord et al., 1997, 1998a). Landscape evolution modeling of crater degradation and icy pinnacle formation, however, rests on the assumption that Callisto's crust includes a substantial abundance of $CO_2$ ice or did so in the geologic past (10% crustal content, White et al., 2016). Consequently, spectral tracers of recently exposed deposits rich in crustal $CO_2$ ice might be present on the ancient surface of Callisto.

If both complexed $CO_2$ and crystalline $CO_2$ ice are present, they could express a double-lobed $CO_2$ feature, with bands centered near 4.25 and 4.27 μm, as is the case on Europa (Villanueva et al. 2023a, Trumbo and Brown 2023). However, NIMS' coarse resolving power (R ~ 40 to 200 between 0.7 to 5.3 μm; Carlson et al, 1992) was too low to disentangle the spectral signatures of these two features and likely would have convolved them into a single band, possibly explaining the ~4.26-μm feature it detected. Similarly, NIMS was unable to resolve the individual rovibrational emission lines of $CO_2$ gas in Callisto's atmosphere, instead detecting a broad, two-lobed peak spanning 4.2 to 4.3 μm (Carlson, 1999). Analyses of absorption bands detected by NIMS were also limited by its low sensitivity at wavelengths $\gtrsim$ 4 μm and the numerous ~0.125 μm-wide filter junctions between 3 and 5 μm that fully or partly overlapped several spectral features of interest, including subtle features between 3.3 and 3.5 μm, the broad 3.88-μm and 4.02-μm bands, and another subtle band near 4.36 μm that might result from the heavy stable isotope $^{13}CO_2$ (McCord et al., 1998a).

Confirming the spectral features detected by NIMS with higher spectral resolution, ground-based observations (R ~ 2500) has proven challenging due to absorption by Earth's atmosphere. Such observations have confirmed the presence of the wide 4.02-μm and 4.57-μm bands (Cartwright et al., 2020), but strong absorption by telluric $CO_2$ has prevented analysis of solid-state and gaseous $CO_2$ features between 4.2 and 4.4 μm. Spectral lines for telluric $CH_4$ and other gases overprint the wavelength range of the subtle bands detected by NIMS between 3.3 and 3.5 μm, complicating their analysis.

The NIRSpec spectrograph on the James Webb Space Telescope (Gardner et al., 2023) is uniquely capable of investigating $CO_2$ and possible organic features, as demonstrated by the recent detection of a double-lobed $CO_2$ feature on Europa (Villanueva et al. 2023a) and high sensitivity characterization of $CO_2$ on Ganymede (Bockelée-Morvan et al., 2023). Here we report JWST/NIRSpec spectral cubes (G395H) of Callisto (Figure 1). These data reveal Callisto's spectral properties at dramatically higher spectral resolution and signal-to-noise ratios (S/N) compared to existing NIMS or ground-based datasets. We used these NIRSpec cubes to measure the spectral properties and spatial distribution (~320 km/spaxel) of solid-state and gaseous $CO_2$ and investigate a suite of other spectral features, some of which are reported here for the first time, including absorption bands that might result from CO and carbonyl sulfide (OCS). Our results shed new light on the evolution of Callisto's surface geology and composition. Our findings also highlight JWST's capabilities for analyzing the spectral properties of $CO_2$, an important molecule in the chemical cycles operating on the Galilean moons and other icy bodies observed across the outer Solar System (*e.g.*, Villanueva et al., 2023a; Trumbo and Brown, 2023; Bockelée-Morvan et al., 2023; de Prá et al., 2023; Brown and Fraser, 2023; Pinto et al, 2023; Emery et al., 2023; Protopapa et al., 2023; Markwardt et al., 2023; Wong et al., 2023a,b).

## 2. Data and Methods

### *2.1 Observations and data reduction*

Callisto was observed with the G395H grating of the NIRSpec spectrograph (2.85 – 5.35 µm, R ~ 2700) on JWST as part of the General Observer Program 2060 (Cartwright et al., 2021). These observations occurred on November 15th and 25th 2022 when the sub-observer point was near longitude 279°W (trailing hemisphere) and 137°W (leading hemisphere, roughly centered on the Asgard impact basin), respectively. Data were collected using NIRSpec's integral field unit (IFU) that has a 3 x 3 arcsecond field-of-view (FOV), placing roughly 165 spaxels (0.1 x 0.1 arcsecond dimensions) across Callisto's disk (Figure 1). Each observation had a total exposure time of ~128 s, spread over four dithers (~32 s each) that sample different parts of the detector, using the NRSRAPID readout mode (see Jakobsen et al. (2022) for more details). G395H data have a ~0.1 µm wide 'unrecoverable' gap that shifts between 4 and 4.2 µm across each of the 29 image slices comprising NIRSpec's IFU. For extended targets like Callisto, the wavelength range of the gap therefore changes across its resolved disk. Consequently, some of the wavelength range covered by the gap can be recovered by extracting spectra from smaller subsets of spaxels on its disk, effectively shrinking the gap in these Callisto data (~4.06 – 4.13 µm).

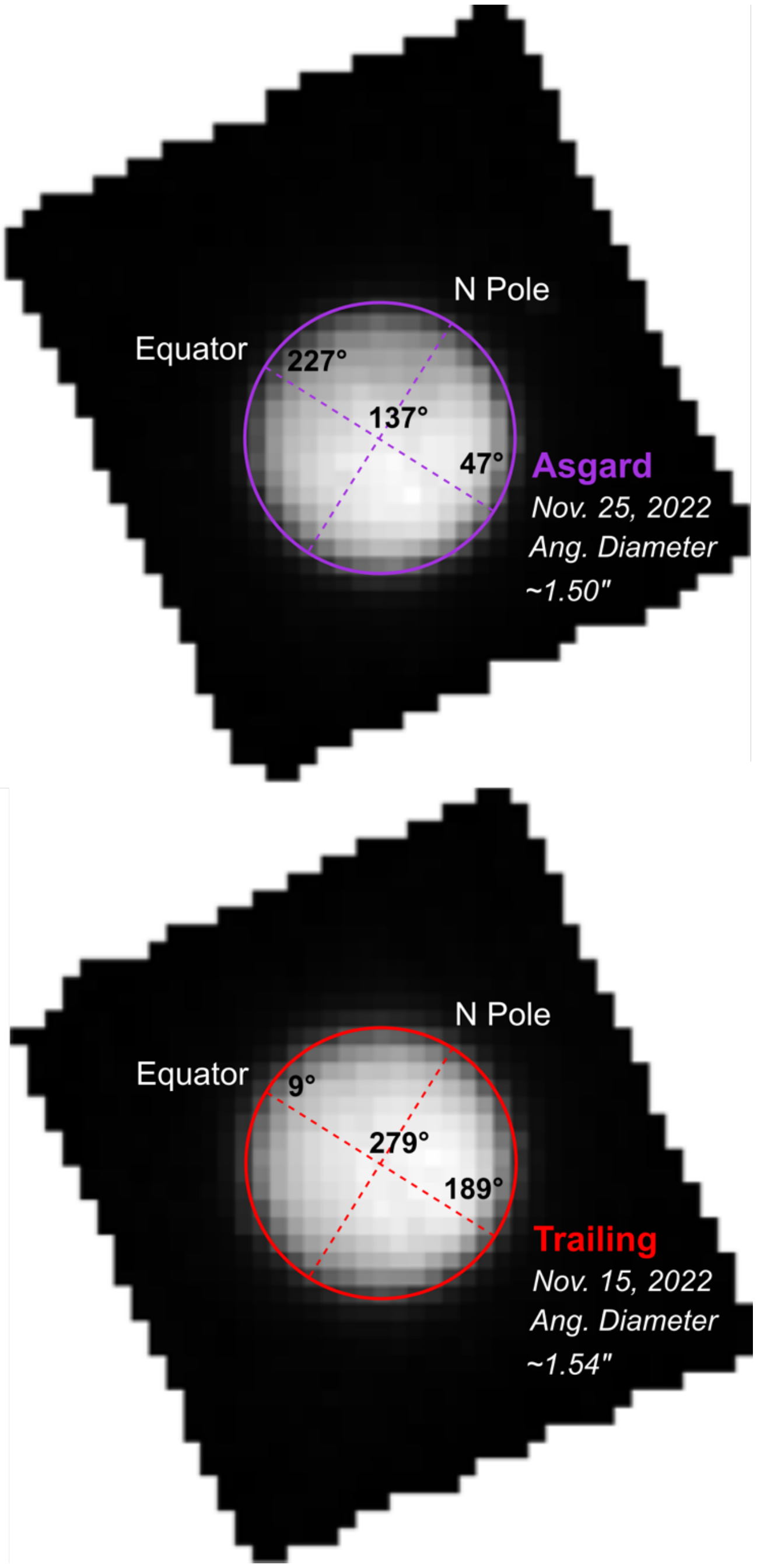

***Figure 1:*** *Raw JWST/NIRSpec spectral cubes (G395H) of Callisto. Dashed lines show the approximate location of Callisto's equator and poles. Sub-observer longitudes for the center and limbs of each cube are labeled.*

All data were downloaded from the Mikulski Archive for Space Telescopes (MAST) at the Space Telescope Science Institute, and they can be accessed via https://doi.org/10.17909/w8qj-5v21. These data were processed using the JWST Science Calibration Pipeline v1.9.4 with CRDS context jwst_1041.pmap (Bushouse et al., 2023). Additional custom codes were developed to combine dithered frames and remove bad pixels (scripts available at github.com/nasapsg). The four dithers were georeferenced to Callisto's disk and then median combined, which assisted in removing abnormal pixels. At each spaxel, we separated the reflectance and thermal signatures of the emission by fitting a two-component model consisting of a realistic solar model and a Planck source function for the thermal radiation. The solar model was generated with the Planetary Spectrum Generator (PSG, Villanueva et al., 2018, 2022). PSG accounts for all Doppler shifts and uses the high-quality ACE solar spectrum (*e.g.*, Hase et al. 2010) to integrate the solar Fraunhofer lines and adopts the Kurucz (2005) solar model to replicate the continuum intensity. After removing the thermal component, reflectance spectra at each spaxel were determined by dividing the calibrated fluxes with a solar model scaled by the projected spatial size of the pixel and corrected for the distances between the Sun and Callisto, and JWST and Callisto, at the time of each observation. A similar process was used to analyze NIRSpec IFU cubes of the Galilean moon Europa (Villanueva et al., 2023a) and the Saturnian moon Enceladus (Villanueva et al., 2023b). Finally, spaxels covering Callisto's disk were summed to generate disk-integrated spectra for the trailing and Asgard observations, hereon referred to as the 'trailing' and 'leading' hemisphere integrated spectra, respectively. Uncertainties for these integrated spectra were estimated by standard error propagation routines that utilize the underlying calibrated uncertainties for each spaxel (as reported by the pipeline).

### *2.2 Band parameter measurements for the integrated spectra*

We identified eleven absorption features for analysis in the integrated spectra of Callisto's trailing and leading hemisphere (Table 1). We measured the band area and depth for each of these eleven features with a band parameter program used previously to measure absorption features in icy satellite spectra (*e.g.*, Cartwright et al., 2022, 2023). The program identifies the continuum within 0.005 to 0.01 μm on both sides of each band and fit it with a line, then divides each band by its continuum. The resulting continuum-divided bands were visually inspected prior to measuring the area and depth of each feature. The program measured the depth of each continuum-divided band by calculating the mean reflectance within ±0.002-0.003 μm of each band center (Table 1) and propagating errors. The band center reflectances were then subtracted from 1 to calculate the band depth for each feature. The program used the trapezoidal rule to calculate band areas and ran Monte Carlo simulations to estimate the 1σ uncertainties by resampling the errors of all channels within the wavelength range of each absorption band.

### *2.3 Spatially resolved band parameter maps*

To investigate the spatial distribution of different species, we generated spectral maps using the individual spaxels in the two thermally-corrected, dither-averaged cubes. We focused our analysis on the three non-$H_2O$ ice features with the strongest bands, centered near 4.25 μm, 4.38 μm, and 4.57 μm, previously attributed to $^{12}CO_2$ (Carlson et al., 1996), $^{13}CO_2$ (McCord et al., 1998a), and other C-bearing species (McCord et al., 1997; Johnson et al., 2004; Cartwright et al., 2020), respectively. We generated continuum-divided band depth and band center maps for these three features. The band-fitting procedure was conducted with the Python *lmfit* package (10.5281/zenodo.11813). We fit a line to the continuum of each band in each spaxel, and then divided by the modeled continuum. After manually inspecting the quality of the resulting continuum-divided bands in each spaxel, the program determined their central wavelength positions, using gaussian fits to each band. To estimate 1$\sigma$ uncertainties for the band depth and center measurements, we used a least squares minimization approach (error maps shown in Figures A1 – A3).

For the 4.25-µm $^{12}CO_2$ band, we utilized a two-gaussian approach to better capture subtle shifts in the band center. For the 4.38-µm $^{13}CO_2$ band, we used a similar two-gaussian fit, where one gaussian fit the entire width of the band (4.34 – 4.42 µm) and another gaussian fit a narrower and deeper feature centered near 4.38 µm, which overprints the broad band in a large number of (but not all) spaxels. Consequently, we mapped the distribution of the 4.38-µm $^{13}CO_2$ band both with and without the narrow 'peak' feature (band depth map for the $^{13}CO_2$ peak feature is shown in Figure A4). The 4.57-µm bands only required single-gaussian fits to conduct satisfactory band depth and band center measurements.

Once the band parameter measurements were finalized, we projected the data onto a regularly spaced latitude/longitude grid. To do this we determined the center of Callisto's disk in the dither-averaged cubes and registered this central point to a specific latitude and longitude, based on the approximate angular radius and north pole position angle of Callisto at the mid-observation time. The latitude and longitude coordinates of each spaxel vertex were then used to form a projectable polygon, using the Python *shapely* package (10.5281/zenodo.7583915). The retrieved measurement parameters and errors for each spaxel, and their map-projected polygons, were stored in separate arrays within a GeoPanda data series (10.5281/zenodo.5573592) for each of the three features we mapped and report here (Figure 2). Because of higher noise and poor geometric sampling in spaxels near the edges of Callisto's disk, we only used spaxels within 0.57" of Callisto's disk center (red polygons in Figure 2) and omitted spaxels beyond this threshold from the finalized spectral maps.

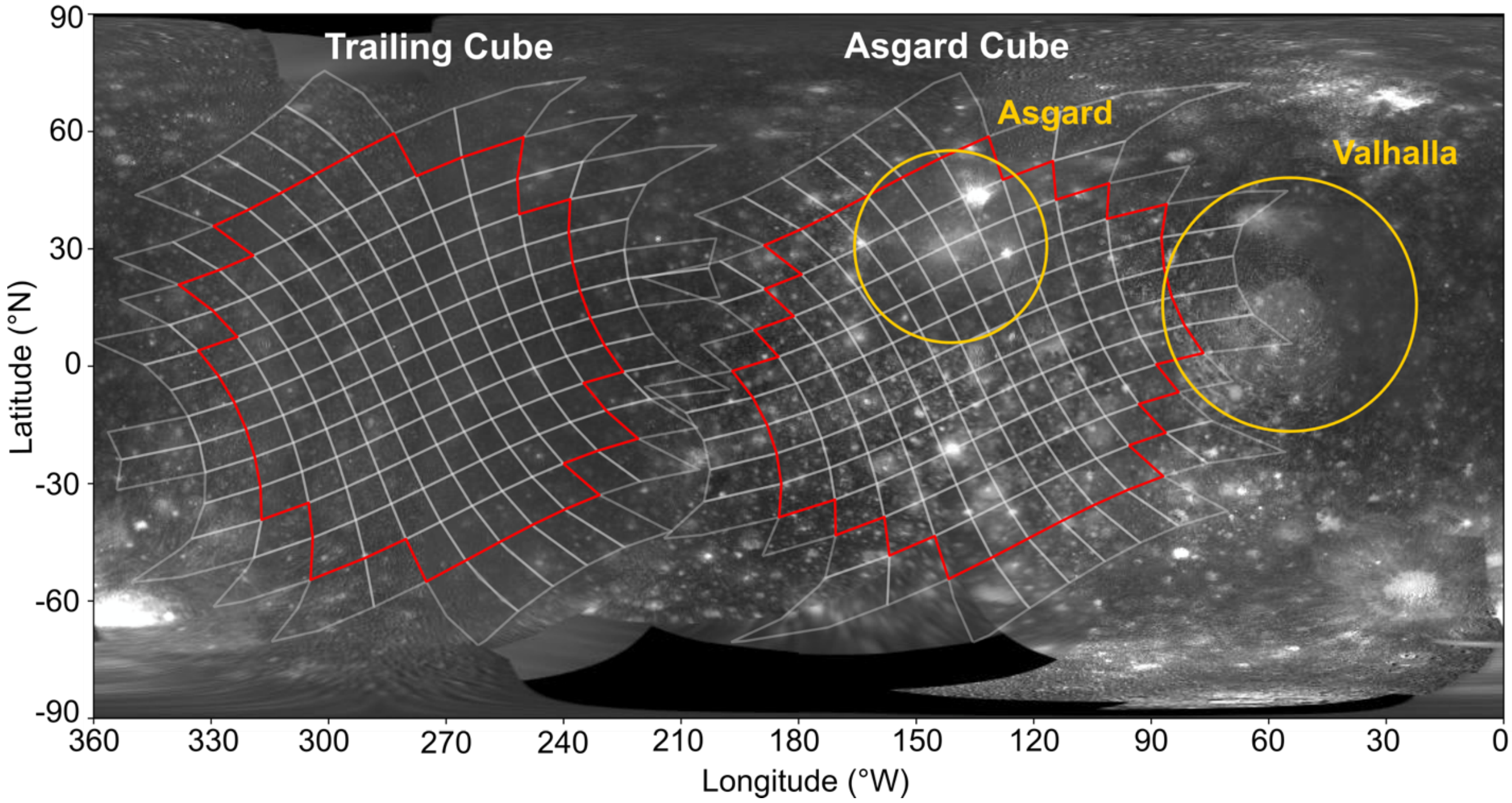

***Figure 2:*** *Callisto base map generated using Galileo Solid State Imager data (NASA/JPL-Caltech/USGS), overprinted by map-projected versions of the full NIRSpec footprints (white polygons) and the higher signal NIRSpec footprints utilized in the results and analyses reported here (red polygons). The approximate boundaries of the Asgard and Valhalla impact basins are indicated (gold circles).*

*2.4 Extraction of $CO_2$ gas emission lines and calculation of column densities*

Atmospheric species like $CO_2$ gas absorb solar radiation, become excited, and then exhibit fluorescent emission. In the case of the strong fundamental $\nu_3$ band of $CO_2$, 'solar-pumped' fluorescence leads to emissions between 4.2 and 4.3 µm in the $\nu_3$ fundamental stretching band, separated into two sets of narrow rovibrational lines, called 'P' and 'R' branches. Although a double-lobed emission peak for $CO_2$ gas was detected by NIMS in Callisto's atmosphere (Carlson, 1999), the individual rovibrational lines could not be detected with NIMS due to its low resolving power (R 40 – 200). The G395H grating (R ~ 3000 at 4.25 µm) is ideally suited for investigating $CO_2$ gas emission lines.

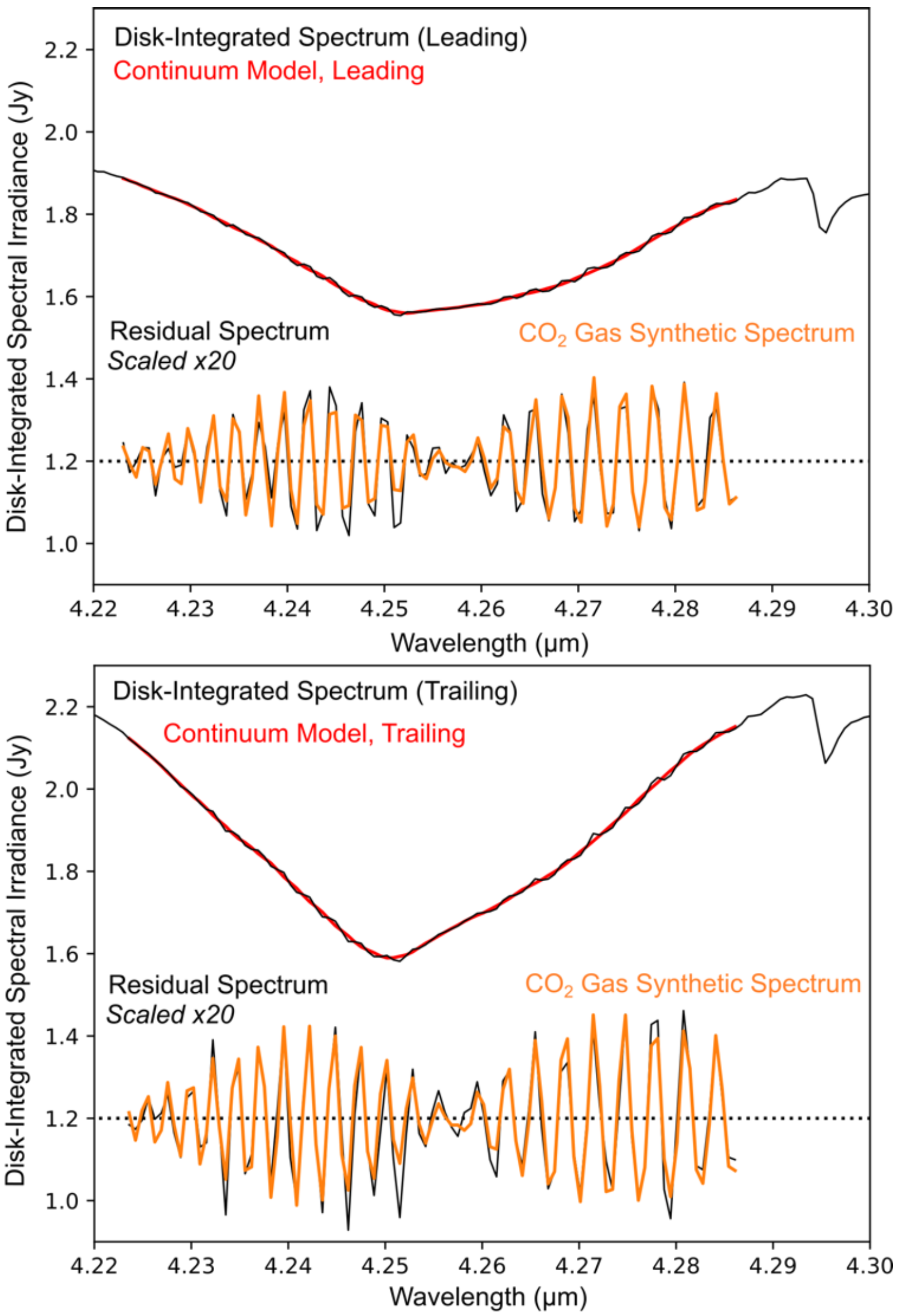

***Figure 3:*** *Integrated NIRSpec spectra for Callisto's leading (top) and trailing (bottom) hemispheres at their native resolutions (black) and continuum models (red). The residual spectra (i.e., spectrum minus continuum model) are plotted below each integrated spectrum, offset from 1.0 and scaled by a factor of 20 for clarity. Best fit $CO_2$ gas synthetic spectra (orange) are plotted over the residual spectra.*

To measure $CO_2$ gas in fluorescence, we first generated a continuum model by smoothing the Callisto spectra between 4.2 and 4.3 µm until the 'sawtooth' pattern detected by NIRSpec was removed from the data (R ~ 1000, Figure 3). We then subtracted the continuum model from the native resolution NIRSpec data (R ~ 3000), generating residual spectra. We performed this technique on all spaxels covering Callisto's disk and a ~0.3" wide annulus of spaxels beyond its disk to search for $CO_2$ gas over a range of altitudes above its surface (~1000 km). Next, we generated synthetic spectra of $CO_2$ gas rovibrational lines between 4.2 and 4.3 µm using PSG, performed the same smoothing/subtraction as applied to the data, and compared these residual models to the residual spectra using cross-correlation (Villanueva et al., 2018, 2022). We iterated this approach, varying the $CO_2$ concentration, until the model matched the 'spikiness' of the residual spectra (synthetic spectra in Figure 3). This approach provided an estimate of the average line-of-sight $CO_2$ gas column density as seen by the observer and did not correct for the observing geometries (*i.e.*, incidence and emission angles). Furthermore, we assume that the excitation process is dominated by solar pumped fluorescence, and meaningful contributions to molecular excitation by electrons at Callisto would likely lower the $CO_2$ gas column density estimates reported here. An approximation of electron excitation of $CO_2$, utilizing an electron population similar to the one used to interpret ultraviolet emissions (Cunningham et al. 2015, Roth 2021) suggests only minor contributions (<0.1%), supporting solar-pumped fluorescence as the dominant excitation process at Callisto. Root-mean-square and chi-square statistics between the finalized continuum-model and the data were used to define the 1σ uncertainties for the resulting $^{12}CO_2$ column density estimates.

## 3. Results and Analyses

### *3.1 Detected absorption features*

*Prominent C-bearing features:* The integrated spectra of Callisto's leading and trailing hemisphere (Figure 4) show conclusive evidence for the global presence of the 4.25-µm $^{12}CO_2$ absorption band detected previously by NIMS (*e.g.*, Carlson et al., 1996). The integrated NIRSpec data confirm that this feature is significantly stronger on Callisto's trailing hemisphere (Table 1). NIRSpec also confirmed the presence of a broad 4.57-µm absorption feature that was previously detected by NIMS and attributed to refractory CN-bearing organic residues (McCord et al., 1997). However, the exact identity of the 4.57-µm band remains uncertain (d'Hendecourt et al., 1986; Accolla et al., 2018; Gerakines et al., 2022) and carbon suboxide ($C_3O_2$; Johnson et al., 2004) and carbon disulfide ($CS_2$; Cartwright et al., 2020) have also been suggested. Furthermore, we report detection of a band centered near 4.38 µm that likely results from the $\nu_3$ mode of $^{13}CO_2$, as measured in the laboratory for $CO_2$ (*e.g.*, Hansen, 1997), which was recently detected on Europa (Villanueva et al., 2023a) and numerous trans-Neptunian objects (TNOs) (*e.g.*, de Prá et al., 2023) using NIRSpec. A subtle absorption band centered near 4.36 µm was originally noted in NIMS data of Callisto and tentatively attributed to $^{13}CO_2$ (McCord et al., 1998a), but no quantitative analyses were made on the feature at that time. Laboratory experiments demonstrate that $^{13}CO_2$ is frequently detected alongside $^{12}CO_2$ in irradiated substrates composed of $H_2O$ ice and carbonaceous material that includes $^{13}C$ (*e.g.*, Bennett et al., 2010), and the presence of this isotope on Callisto is therefore unsurprising, albeit a similarly broad 4.38-µm features detected in G395H NIRSpec data of Ganymede (Bockelée-Morvan et al., 2023) and Io (de Pater et al., 2023) might result from calibration artifacts. We consider whether Callisto's 4.38-µm feature may be contaminated by an artifact in section 4.3 and Appendix A.3.

*$H_2O$ features:* The integrated spectra show definitive evidence for $H_2O$ via the global detection of a strong 3-µm band, resulting from $H_2O$ ice and hydrated minerals, and a 3.1-µm Fresnel peak, indicative of crystalline $H_2O$ ice (*e.g.*, Grundy and Schmitt, 1998; Mastrapa et al., 2009) (Figure 4). We find no convincing evidence for the broad 4.5-µm $H_2O$ ice feature nor the 3.6-µm $H_2O$ ice peak, consistent with prior analyses of Callisto using NIMS (*e.g.*, McCord et al., 1998a) and ground-based data (Cartwright et al., 2020). These relatively weaker $H_2O$ ice features could be obscured by a global ~1 cm thick layer of dark dust (likely thicker in some locations), based on the analysis of Callisto's radar backscatter properties (Ostro et al., 1992; Black et al., 2001; Moore et al., 2004). Additionally, the absence of these $H_2O$ ice features indicates that hydrated minerals are a significant reservoir of $H_2O$ on Callisto's surface and contribute to its strong 3-µm band, supporting prior assessments of Callisto's $H_2O$ inventory (*e.g.*, McCord et al., 1998a). Unlike Callisto, the 4.5-µm and 3.6-µm $H_2O$ ice features are present in NIRSpec data of Europa (Villanueva et al., 2023a) and Ganymede (Bockelée-Morvan et al., 2023) and have been detected in other datasets of the icy Saturnian moons (*e.g.*, Emery et al., 2005; Cruikshank et al., 2005), Saturn's rings (*e.g.*, Hedman et al., 2023), the Uranian moons (Cartwright et al., 2018), Charon (Protopapa et al., 2023), and other TNOs (*e.g.*, De Prà et al., 2023), highlighting the relatively distinct spectral signature of $H_2O$ on Callisto. Detailed analysis of $H_2O$ features is beyond the scope of this project and left for future work.

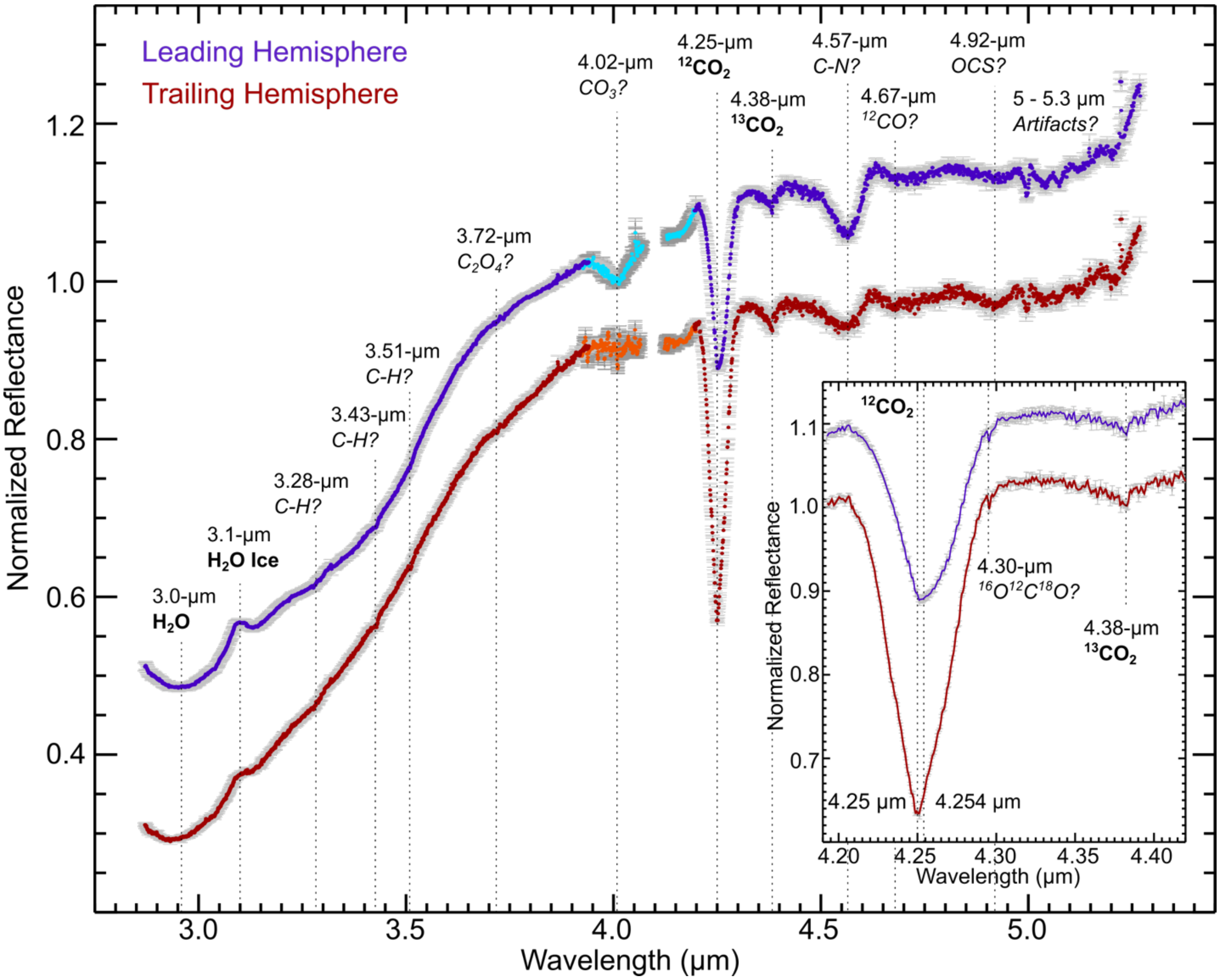

***Figure 4:*** *JWST/NIRSpec integrated spectra and 1σ uncertainties (gray error bars) of Callisto's leading (purple) and trailing (red) hemisphere, normalized to 1 at 3.82 μm and offset vertically for clarity. The G395H grating has a ~0.1 μm wide wavelength gap, shifting between ~4 and 4.2 μm across NIRSpec IFU's 29 image slices. Some of the image slices that span Callisto's disk include wavelength coverage between ~4 and 4.2 μm, which are shown here for Callisto's leading (bright blue) and trailing (bright orange) hemispheres. All spectral features identified in this study are labeled, with dotted lines indicating their central wavelengths. Features with confirmed compositions have bolded labels, whereas weak features, or those with multiple compositional interpretations, are italicized and followed by question marks. Possible bands and spectral structure at wavelength >4.98 μm may result from data calibration artifacts and are not analyzed in this study. The inset box shows a close-up of the 4.19 to 4.42 μm wavelength range, highlighting the $CO_2$ features we have identified and the different band centers for $^{12}CO_2$ on Callisto's leading (4.254 μm) and trailing (4.250 μm) hemisphere, as well as a feature near 4.3 μm that may result from an isotope of $CO_2$ or could be a residual solar line (Appendix A.3).*

*Other detected features:* Six other, more subtle features centered near 3.28, 3.43, 3.51, 3.72, 4.67, and 4.92 µm are also apparent in the integrated spectra (Figure 4). The 4.67-µm and 4.92-µm features have not been previously identified on Callisto, and we report their detection for the first time. Subtle features between 3.3 and 3.5 µm were previously identified in NIMS spectra (McCord et al., 1998a) and some ground-based datasets (Cartwright et al., 2020). Subtle features between 3.7 and 3.8 µm were identified in ground-based data (Cartwright et al., 2020) and some NIMS spectra (Michael J. Malaska, private communication). These subtle absorption features could be associated with the presence of carbon-rich species, in particular $CO_2$ and other carbon oxides, as well as possible CH-bearing (hydrocarbons), CN-bearing (nitriles and isonitriles), and CS-bearing constituents.

The subtle 3.43-µm and 3.51-µm features detected in NIRSpec data of Ganymede may result from calibration artifacts (Bockelée-Morvan et al., 2023). Because 3.4-µm and 3.5-µm bands were previously detected in NIMS (McCord et al., 1998a) and some ground-based (Cartwright et al., 2020) datasets of Callisto, we think these two features are probably real and associated with surface components. Other detected features that may be spurious, or were detected in other datasets but not in the integrated NIRSpec spectra, are described in Appendix A.3. All detected features with band depths $\lesssim$1% of the continuum likely require follow-up observations by JWST and other telescope facilities to corroborate their presence. We consider the species that could be contributing to these subtle features in section 4.5.

*3.2 Band area and depth measurements*

We conducted continuum-divided, band area and depth measurements on the three strongest absorption bands near 4.25, 4.38, and 4.57 µm, finding that they are present at >3σ levels on both hemispheres (Table 1). The prominent 4.25-µm $^{12}CO_2$ band is the most ubiquitous non-$H_2O$ feature on Callisto and is significantly stronger (>>3σ difference) on its trailing hemisphere (band depth ~33%) compared to its leading side (band depth ~19%) (Table 1, Figure 4). The strong 4.57-µm band is also ubiquitous and displays a significant hemispherical asymmetry (>>3σ difference), with a notably stronger band on Callisto's leading side (band depth ~6%) compared to its trailing side (band depth ~3%). The broad 4.38-µm band is somewhat evenly distributed across Callisto's leading and trailing hemispheres and is only marginally stronger on its trailing side (respective band depths of ~2.3% and ~2.6%, <1σ difference). The spatial trends and hemispherical distributions of these three bands are described in greater detail in section 3.4.

We also conducted continuum-divided, band area and depth measurements on the six subtle spectral features centered near 3.28, 3.43, 3.51, 3.72, 4.67, and 4.92 µm, as well as possibly spurious bands near 4.3 µm and between 5 and 5.3 µm (described in Appendix A.3), finding that they are all present at >3σ levels with band depths ranging between 0.6 to 1.6% (Table 1). The 3.72-µm and 4.67-µm features are notably weaker on Callisto's leading side (<3σ detection) and we only report their detection on Callisto's trailing hemisphere. Similarly, the 4.92-µm feature is modestly stronger on Callisto's trailing hemisphere (>1σ difference in band depth, >3σ difference in band area). The implications of these hemispherical asymmetries are considered in greater detail in section 4.1.

*3.3 Estimated column densities and distribution of $CO_2$ gas*

We calculated $CO_2$ gas column densities ranging between ~0.4 to 1.0 x$10^{19}$ $m^{-2}$ on both sides of Callisto. These NIRSpec-derived estimates are consistent with column density estimates for Callisto's leading hemisphere made using NIMS data (0.8 x$10^{19}$ $m^{-2}$; Carlson, 1999). The signature of $CO_2$ gas in the NIRSpec data drops off substantially beyond the edge of Callisto's disk (~320 km/spaxel), suggesting that it is retained close to its surface, consistent with the previous characterization of Callisto's $CO_2$ atmosphere (estimated to be between ~5 to 40 km above its surface; Carlson, 1999).

The column density maps indicate that the distribution of $CO_2$ gas does not coincide with the regions of peak reflected emission and peak surface temperatures (Figure 5). To estimate Callisto's temperature, we fit a two black-body function to its spectrum in each spaxel, consisting of a Planck function set to 5777 K (reflected solar component), and a Planck function where temperature is a free parameter (thermal emission component). On Callisto's leading hemisphere, peak $CO_2$ gas column densities are associated with spaxels slightly east of Valhalla, whereas minimum $CO_2$ column densities are in spaxels clustered on Callisto's anti-Jovian side. Solid-state $CO_2$ exhibits slightly larger band depths in spaxels near the relatively bright Asgard and slightly east of Valhalla compared to the darker surrounding terrains (Figure 6), perhaps contributing to the higher column densities near Valhalla. On Callisto's trailing side, peak $CO_2$ gas column densities are near 290°W and 45°S, notably offset from peak reflected emissions and peak surface temperatures. Additionally, the largest $CO_2$ column densities on Callisto's trailing side are clearly offset from the spaxels that exhibit the strongest solid-state $CO_2$ bands (Figure 6).

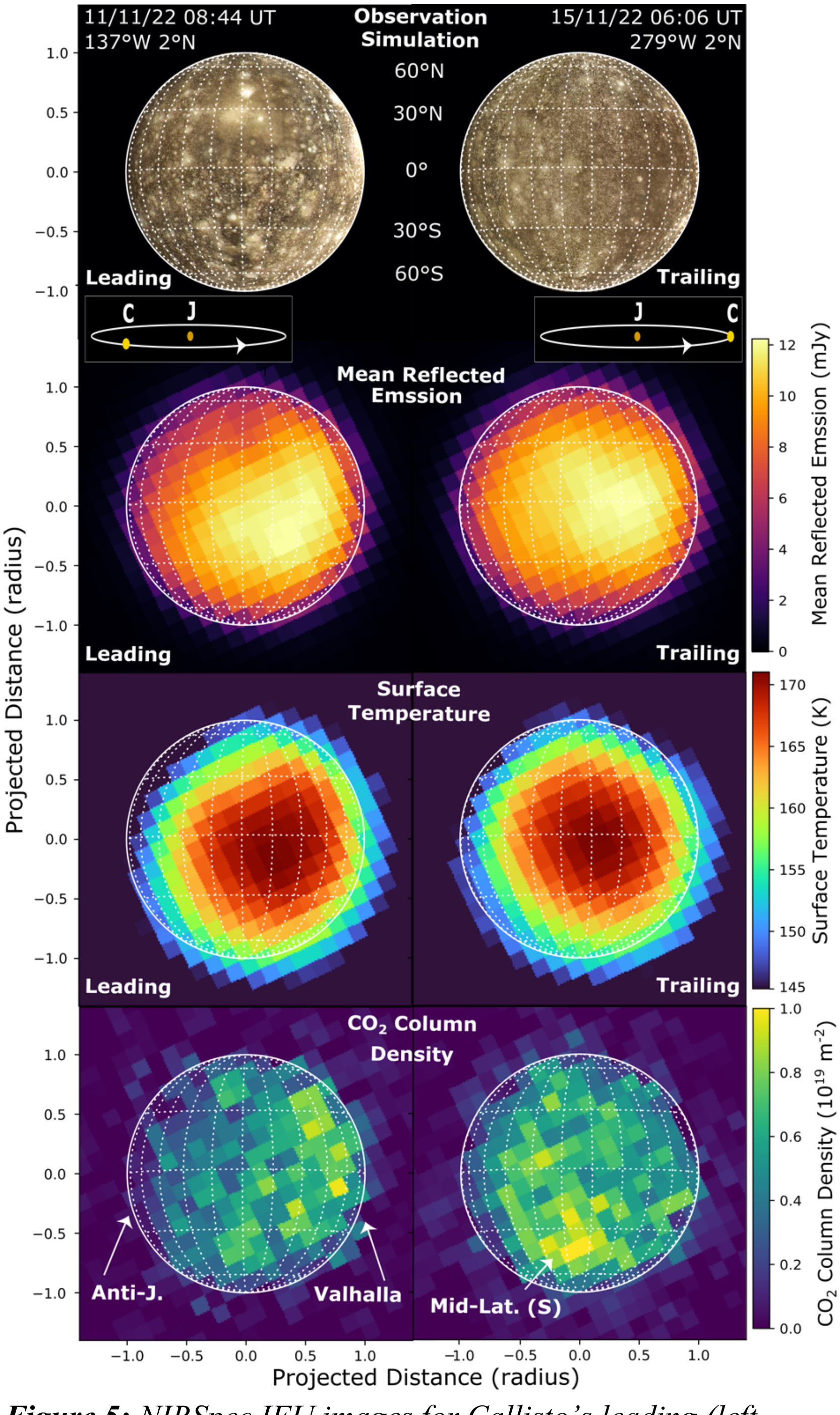

***Figure 5:*** *NIRSpec IFU images for Callisto's leading (left column) and trailing (right column) hemisphere. The globes show the simulated observing geometries (inset plot shows the orbit of Callisto around Jupiter), the mean reflected emission at 2.9 µm, the estimated surface temperatures, and the retrieved $CO_2$ column densities, from the top row to bottom row, respectively. Notable asymmetries in the distribution of $CO_2$ gas are indicated with white arrows (discussed in section 4.2).*

*3.4 Spatial distribution of solid-state $CO_2$ and the 4.57-μm band*

*$^{12}CO_2$:* Consistent with the integrated spectra, our spectral maps highlight the ubiquitous presence of $CO_2$ on Callisto (Figure 6, error maps shown in Figure A1). Near the center of Callisto's trailing hemisphere, continuum-divided band depths range between 35 to 40%, representing the largest values we measured. The depth of the $^{12}CO_2$ feature gradually decreases away from this central point, toward transitional longitudes (180° and 360°), ranging between 30 to 35% at low latitudes (30°S – 30°N) and 25 to 30% at mid latitudes (30 – 60°). This distribution of $^{12}CO_2$ mirrors the bullseye pattern identified in the NIMS dataset, where peak band depths at low latitudes near the center of Callisto's trailing side are up to 40% of the continuum (see Plate 4 in Hibbitts et al., 2000). The band depths of the $^{12}CO_2$ feature on Callisto's leading side are lower than on its trailing side (Figure 6), consistent with the integrated spectra and prior measurements made with NIMS. In the NIRSpec cubes, the spaxels covering the anti-Jovian side of Asgard and the sub-Jovian side near Valhalla display the largest band depths measured on the leading hemisphere (20 to 25%). In between these zones, near the center of Callisto's leading side (sub-observer longitude 90°), band depths are notably lower (18 – 20%). In the higher spatial resolution NIMS dataset, $^{12}CO_2$ band depths can approach 40% of the continuum in craters and their ejecta blankets on Callisto's leading side, but outside of these features, $^{12}CO_2$ band depths are much lower, ranging between ~5 to 20% of the continuum (*e.g.*, Figures 5 and 6 in Hibbitts et al., 2002). These $CO_2$-enriched craters and their ejecta blankets are too small to be resolved with NIRSpec. The band center of the $^{12}CO_2$ feature is remarkably consistent across Callisto's trailing hemisphere (4.250 ± 0.002 μm) (Figure 6). On Callisto's leading side, the band center for $^{12}CO_2$ is similar to its trailing side (4.252 ± 0.002 μm), except for the spaxels covering and proximal to Asgard and Valhalla, where the band center is notably shifted to longer wavelengths (4.258 ± 0.002 μm).

*$^{13}CO_2$:* The heavy stable isotope $^{13}CO_2$ is present across Callisto (Figure 7, error maps shown in Figure A2), and it exhibits much weaker continuum-divided band depths (1 – 5%) compared to the $^{12}CO_2$ feature (18 – 40%). The morphology of the $^{13}CO_2$ feature in the integrated spectra is similar on Callisto's leading and trailing sides. At the individual spaxel scale, however, we have identified two components to the $^{13}CO_2$ band, with a broad 'base' feature (spanning 4.335 to 4.415 μm) exhibiting smaller band depths (1 – 3%) overprinted by a narrow 'peak' feature centered close to 4.38 μm that shows larger band depths (2 – 5%). Both the base and peak components of the $^{13}CO_2$ feature are stronger near the center of Callisto's trailing hemisphere and are generally weaker on Callisto's leading side, similar to the distribution of $^{12}CO_2$ (Figures 7 and A4). The band center of the peak feature is essentially unchanging and very near 4.38 μm. To investigate possible wavelength shifts in the broader base component, we utilized single-gaussian model fits that ignore the narrow peak. The resulting map shows that the base component is somewhat shifted to shorter wavelengths in spaxels on Callisto's trailing side compared to its leading side, but the overall distribution is fairly mottled (Figure 7). In general, regional variations in $^{13}CO_2$ band depths and centers are more ambiguous than the same measurements of $^{12}CO_2$ or the 4.57-μm band (see below), likely because $^{13}CO_2$ bands are much weaker, making spatial trends hard to discern.

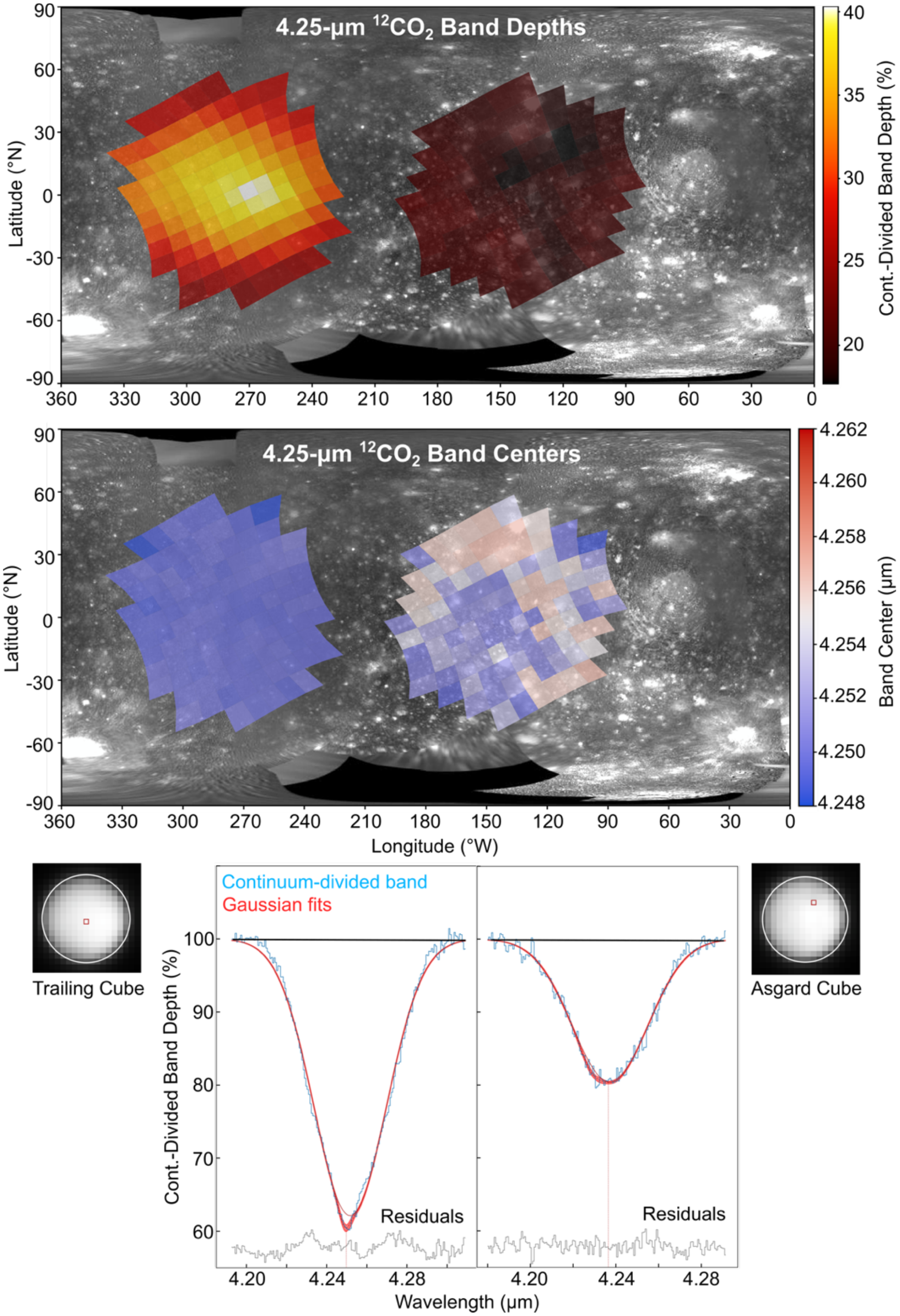

***Figure 6:*** *4.25-µm $^{12}CO_2$ band depth (top) and band center (middle) maps. Example continuum-divided 4.25-µm bands and model fits (bottom) are shown for spaxels near the center of Callisto's trailing side (left) and in the Asgard impact basin (right). These maps highlight the substantially stronger $^{12}CO_2$ feature on Callisto's trailing side and the shifted central wavelength of the $^{12}CO_2$ band on Callisto's leading side.*

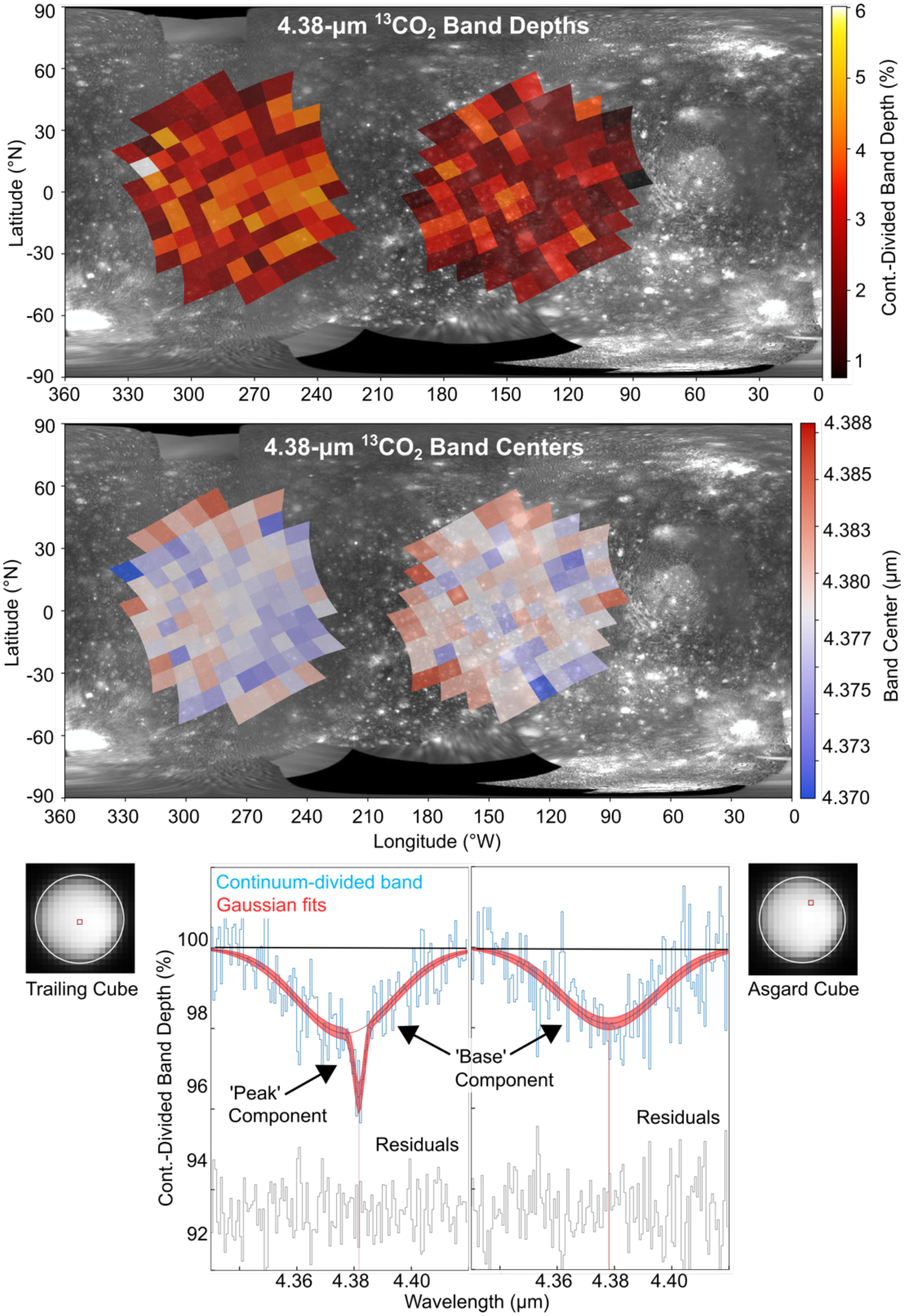

***Figure 7:*** *4.38-µm* $^{13}CO_2$ *band depth map, including both the 'base' and 'peak' components of this feature (top) and 4.38-µm band center map for the base component only (middle) (the peak component does not exhibit wavelength shifts and so is excluded). Example continuum-divided 4.38-µm bands and model fits (bottom) are shown for spaxels near the center of Callisto's trailing side (left) and in the Asgard impact basin (right). These maps highlight the modestly stronger* $^{13}CO_2$ *feature on Callisto's trailing side, in part resulting from the narrow peak component overprinting the wide and shallow base component. This narrow peak feature is fit by a second gaussian model (shown in bottom, left plot).*

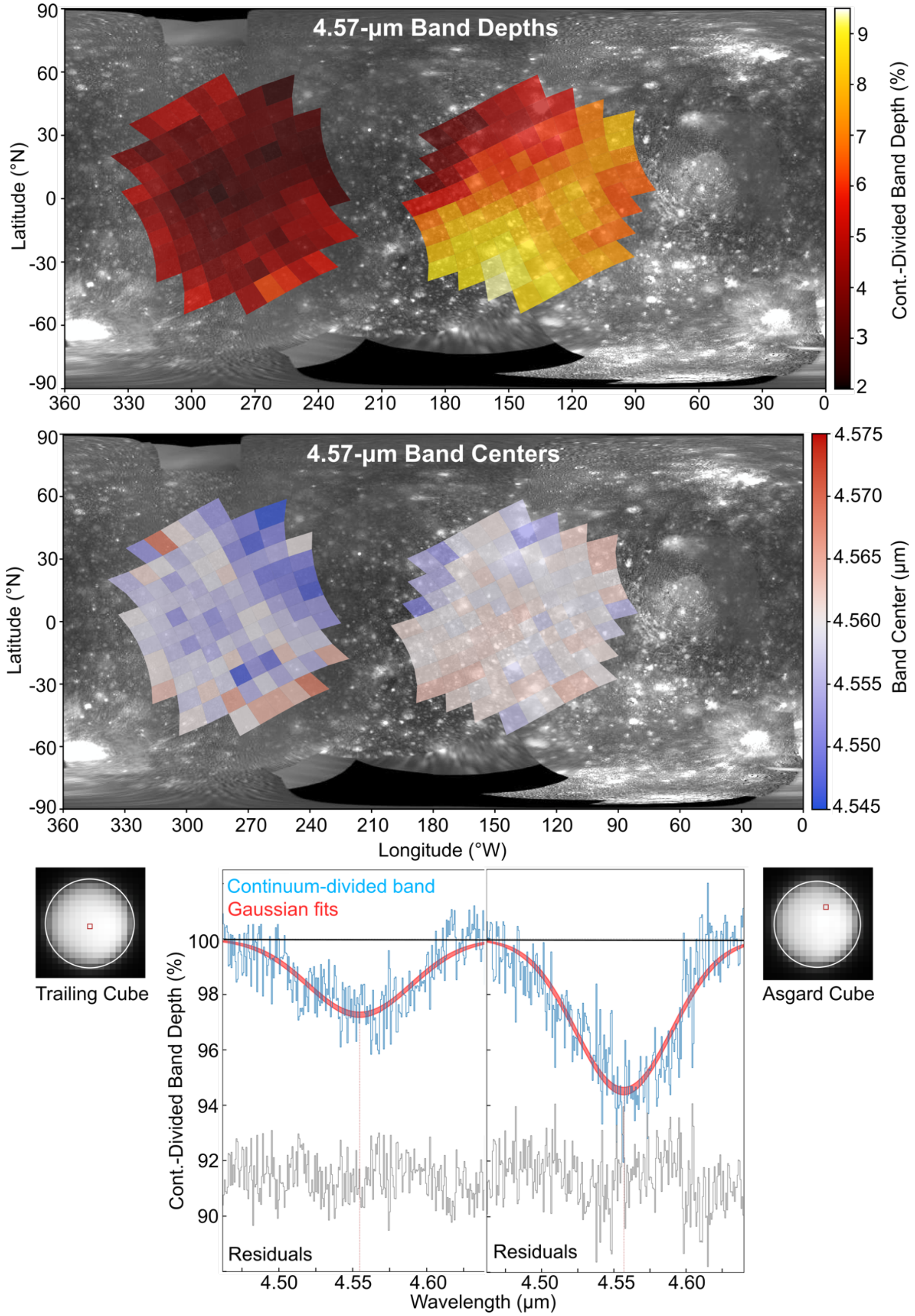

***Figure 8:*** *4.57-µm band depth (top) and band center (middle) maps. Example continuum-divided 4.57-µm bands and model fits (bottom) are shown for spaxels near the center of Callisto's trailing side (left) and in the Asgard impact basin (right). These maps highlight the notably stronger 4.57-µm feature on Callisto's leading side, and the shift of its band center to shorter wavelengths on Callisto's trailing side.*

*4.57-μm feature:* The continuum-divided band depth map for the 4.57-μm band (Figure 8, error maps shown in Figure A3) shows that this feature is weaker on Callisto's trailing hemisphere (2 – 6%) compared to its leading side (2 – 10%). The smaller band depths are most obvious near the center of Callisto's trailing hemisphere (2 – 4%). The 4.57-μm band depths show prominent regional variations on Callisto's leading hemisphere, with the largest band depths in terrains outside of Asgard, especially at mid southern latitudes (8 – 10%). Within Asgard, and at low and mid northern latitudes toward Callisto's anti-Jovian side, 4.57-μm band depths are lower (2 – 6%). The band centers for the 4.57-μm band range between 4.54 and 4.58 μm on both hemispheres (Figure 8), but the average band center appears to be shifted to shorter wavelengths (4.555 ± 0.010 μm) on Callisto's trailing side compared to its leading side (4.565 ± 0.010 μm), although these wavelength ranges overlap. Thus, spaxels that exhibit larger 4.57-μm band depths tend to have band centers at slightly longer wavelengths.

## 4. Discussion

### *4.1 Origin and nature of solid-state $CO_2$*

We measured the spectral signatures of $^{12}CO_2$ and $^{13}CO_2$ across Callisto's surface. The strong hemispherical dichotomy in the strength of the $^{12}CO_2$ band, and its shifted central wavelength between Callisto's trailing (~4.250 μm) and leading (~4.252 μm) hemispheres compared to Asgard and Valhalla (~4.258 μm), suggests that the origin of $CO_2$ could involve more than one source and/or $CO_2$ is modified by different processes in different locations. The bullseye shaped distribution of $^{12}CO_2$ on Callisto's trailing side is consistent with bombardment by fast rotating plasma in the Jovian magnetosphere that primarily interacts with the trailing sides of the Galilean moons. In this scenario, carbonaceous material mixed with $H_2O$ is irradiated, forming $CO_2$ molecules and other carbon oxides. The radiolytically generated $CO_2$ molecules are bound or trapped in the host regolith materials (Hand and Carlson, 2012), allowing $CO_2$ to build up over time. This trapping process could help explain how this volatile is able to persist at Callisto's peak surface temperatures (~170 K, Figure 5). Additionally, the subtle bands centered near 3.72, 4.30, 4.67, and 4.92 μm might result from various oxides of carbon (section 4.5). The presence of these features on Callisto's trailing hemisphere, and their relative weakness or absence from Callisto's leading side (Table 1), also supports radiolytic production and efficient trapping of carbon oxides in dark material and/or $H_2O$ ice on Callisto's trailing side.

The more regionally variable distribution of $^{12}CO_2$ on Callisto's leading hemisphere implicates additional, non-radiolytic processes. Based on the analysis of NIMS data, it has been suggested that $CO_2$ might be sourced from Callisto's crust and exposed in fresher craters associated with Asgard and Valhalla (Hibbitts et al., 2002). Although the NIRSpec data have insufficient spatial resolution to discern the spectral properties of individual craters, the data do show that $CO_2$ band depths are greater in spaxels covering and proximal to Asgard and Valhalla compared to the surrounding terrains (Figure 6). Crustal deposits could include crystalline $CO_2$ ice, which should rapidly sublimate at Callisto's peak surface temperatures, as suggested by landscape evolution models (White et al., 2016). Perhaps small amounts of residual $CO_2$ ice are present in Asgard and Valhalla, shifting the complexed $CO_2$ feature to longer wavelengths (~4.258 μm), but at insufficient levels to express a second feature at 4.27 μm, unlike on Europa (Villanueva et al., 2023a; Trumbo and Brown, 2023). Another possibility is that the $CO_2$ is trapped in $H_2O$ ice, which has been suggested to explain the wavelength shifts exhibited by $^{12}CO_2$ in spaxels associated with Ganymede's north polar region (Bockelée-Morvan et al., 2023), albeit the spectral signature of $H_2O$ ice is much weaker on Callisto than on Ganymede.

Alternatively, perhaps the $CO_2$ on Callisto's leading hemisphere is primarily generated by radiolysis on its trailing side, which then gets sputtered or sublimates during peak dayside temperatures, migrating in Callisto's tenuous atmosphere to the nightside leading hemisphere, where it subsequently condenses on bright crater rims and rings in Asgard and Valhalla. Such a cold trapping process might form deposits that include crystalline $CO_2$ ice, or alternatively, $CO_2$ could get trapped by co-condensing $H_2O$, thereby explaining the wavelength shift in the $^{12}CO_2$ feature exhibited by Asgard and Valhalla.

The $^{13}CO_2$ band is stronger near the center of Callisto's trailing side, coincident with the strongest $^{12}CO_2$ band depth measurements, suggesting that $^{13}CO_2$ is also generated by radiolysis. Whether the origin of the narrow peak and broad base components of the $^{13}CO_2$ band are linked is uncertain. One possibility is that the narrow peak centered near 4.38 μm represents 'pure' $^{13}CO_2$, whereas the broad base component represents $^{13}CO_2$ mixed with $H_2O$ or other species, possibly explaining the band center shifting (~4.37 to 4.38 μm) exhibited by the base component on Callisto's trailing side. Furthermore, both $^{12}CO_2$ and $^{13}CO_2$ features exhibit shifts to shorter wavelengths on Callisto's trailing hemisphere (Figures 7 and 8, respectively), consistent with complexation on its trailing side and at least some 'free' deposition on its leading side (*i.e.*, formation of deposits where $CO_2$ molecules are primarily surrounded by other $CO_2$ molecules). The 4.57-μm band exhibits a similar wavelength shift on Callisto's leading vs. trailing hemisphere, hinting at similar differences in deposition for the species contributing to this feature (Figure 8). Future laboratory studies that investigate irradiation of $H_2O$ mixed with carbonaceous components under conditions relevant to Callisto are likely required to gain better understanding of the observed wavelength shifts. Of note, the base component of Callisto's 4.38-μm band may be spuriously enhanced by a calibration artifact, as has been suggested for Ganymede 4.38-μm band (Bockelée-Morvan et al., 2023). We discuss this possibility in section 4.3 and Appendix A.3. Additionally, improvements to the NIRSpec data calibration pipeline could help resolve this ambiguity.

### *4.2 A localized and patchy $CO_2$ atmosphere*

The $CO_2$ column densities on Callisto's leading and trailing sides show notable asymmetries in the distribution of $CO_2$ gas that are offset from the subsolar region (Figure 5). The higher $CO_2$ gas abundances in spaxels slightly east of Valhalla could result from sublimation of modestly larger abundances in solid-state $CO_2$, as implicated by the stronger $CO_2$ bands in this location compared to the surrounding terrains (Figure 6). Thermophysical modeling of thermal observations made with the Atacama Large Millimeter/submillimeter Array (ALMA) at 0.87 mm/343 GHz suggests there is an anomalously warm region east and south of Valhalla (Camarca et al., 2023). This warmer region might increase the mobility of solid-state $CO_2$ in Callisto's near-surface, perhaps increasing the rate at which it diffuses out of the regolith, thereby increasing $CO_2$ gas column densities over this location. Sputtering of $CO_2$ molecules into the exosphere (*e.g.*, Raut and Baragiola, 2013) could contribute to $CO_2$ gas abundances over Valhalla, perhaps enhanced by draping of the Jovian magnetic field lines, increasing high energy particle bombardment in some locations (Liuzzo et al., 2019). Geologic processes, such as outgassing of $CO_2$ from Callisto's crust and increased sublimation of $CO_2$ from more recent exposure of crustal deposits of $CO_2$ ice could also contribute to the enhancement in $CO_2$ gas above the Valhalla region.

We report the first detection of $CO_2$ gas over the trailing hemisphere, confirming that $CO_2$ is globally present in Callisto's atmosphere. The peak $CO_2$ column densities over the trailing side are notably offset (~45°S) from the low latitude zone where the estimated surface temperatures and reflected emissions are highest (Figure 5). This location is also offset from the location of the strongest solid-state $CO_2$ absorption bands (Figure 6) (section 4.1). Thus, peak $CO_2$ gas abundances do not coincide with the region where sublimation, sputtering, and radiolysis should be most pronounced on Callisto's trailing side, suggesting that geologic terrains help sustain the $CO_2$ atmosphere above this location. A large 'light plains' unit (*lp*; Bender et al., 1997) centered near 290°W and spanning ~15°S to 45°S (Figure 17.33 in Moore et al., 2004) is slightly north of the peak $CO_2$ column densities, possibly contributing to them. These *lp* units represent impact structures with higher albedos and lower crater frequencies compared to the surrounding terrains (Bender et al., 1997). Whether a cluster of brighter impact structures would spur outgassing of $CO_2$ from Callisto's crust or enable more efficient sublimation or sputtering of solid-state $CO_2$ in this location is uncertain, and the rough spatial association between this *lp* unit and peak $CO_2$ gas column densities could be coincidental. Nevertheless, our results suggest that geologic sources of $CO_2$ are important for sustaining Callisto's atmospheric $CO_2$.

The derived column densities over both hemispheres suggest that $CO_2$ gas is less abundant than atmospheric $O_2$, for which average column densities of 2-4 $\times10^{19}$ $m^{-2}$ were inferred from observations (Cunningham et al. 2015, Hartkorn et al. 2017, de Kleer et al. 2023), but exceeds the abundance of O (Cunningham et al. 2015) and H (Roth et al., 2017). Abundances of $H_2$ and $H_2O$, which are likely present in Callisto's atmosphere (*e.g.*, Carberry Mogan et al. 2022), could be similar to the $CO_2$ densities derived here, suggesting a mixed atmosphere with different species dominating at different locations.

*4.3 $^{13}CO_2/^{12}CO_2$ isotopic ratios*

Ratios between stable isotopes of different elements, such as $^{12}C$ and $^{13}C$, can provide important information about the formation conditions of different icy bodies, the materials they accreted, and the processes that may have subsequently altered these materials. Prior studies have utilized isotopic ratios of remotely sensed spectral features on planetary bodies to gain insight into isotopic ratios of their constitutive elements (*e.g.*, Clark et al., 2019; Grundy et al., 2023; Glein et al., 2023). For example, $^{13}C/^{12}C$ ratios derived from equivalent width measurements of solid-state $^{13}CO_2$ and $^{12}CO_2$ on Saturn's native moon Iapetus are consistent with 'terrestrial' values, exhibited by the inner planets, main belt asteroids, and Saturn's rings (Clark et al., 2019). In contrast, the same measurement technique applied to Saturn's captured moon Phoebe yields a $^{13}C/^{12}C$ ratio enhanced by a factor of ~5 relative to terrestrial values (Clark et al., 2019). This enhancement might arise because Phoebe, which likely formed in the primordial Kuiper Belt and was captured by Saturn (*e.g.*, Johnson and Lunine, 2005), accreted material from a region of the protoplanetary disk where $^{12}C$-bearing gas was shielded from UV photon processing, allowing preferential accretion of $^{13}C$-rich solids formed from photolysis of $^{13}C$-bearing gas (Neveu et al., 2020).

To provide additional context on Callisto's $CO_2$ features and to investigate the possible sources of carbon on its surface, we calculated hemispherical $^{13}CO_2/^{12}CO_2$ band area and depth ratios (measurements listed in Table 3), using the 4.25-µm and 4.38-µm features identified in the integrated spectra (Figure 4). The resulting ratios are notably larger on Callisto's leading side due to the large hemispherical difference in $^{12}CO_2$ but only modest difference in $^{13}CO_2$ (Table 3, Figure 4). Using the same band measurement program, we calculated $^{13}CO_2/^{12}CO_2$ band area and depth ratios for the Saturnian moons Iapetus and Phoebe, utilizing the same data as prior analyses, collected by Cassini's Visual and Infrared Mapping Spectrometer (VIMS) (Clark et al., 2019). We also resampled the NIRSpec spectra of Callisto to simulate the native resolution of the VIMS data at 4.3 µm (R ~ 130) (Figure 9).

The resulting $^{13}CO_2/^{12}CO_2$ band area ratios for Callisto's leading and trailing hemispheres are comparable to Phoebe (<2$\sigma$ difference) and significantly larger than Iapetus (>4$\sigma$ difference) (Figure 10). In contrast, both sides of Callisto have similar band depth ratios to Iapetus (<1$\sigma$ difference), whereas Phoebe's band depth ratio is notably higher than Iapetus or Callisto (>2$\sigma$ difference) (Figure 10). Thus, Callisto's $^{13}CO_2/^{12}CO_2$ band area ratios suggest that its surface may be enriched in $^{13}C$ similar to Phoebe, but its band depth ratios are more consistent with the terrestrial-like values of $^{13}C$ exhibited by Iapetus (the $^{13}CO_2/^{12}CO_2$ band depth ratios have larger uncertainties than the band area ratios, which likely contributes to the ambiguity).

One possibility to explain this discrepancy is that other species contribute to the wide but shallow base component of Callisto's $^{13}CO_2$ band, increasing its band area ratios but not its band depth ratios, thereby enhancing its band area ratios. Alternatively, it has been suggested that a broad and subtle 4.38-µm feature detected on Ganymede (Bockelée-Morvan et al., 2023) and Io (de Pater et al., 2023) could be a calibration artifact and may not result from $^{13}CO_2$. The detection of this feature on Io is of particular interest because these NIRSpec IFU observations were collected during Jupiter eclipse, when Io's spectral properties should be dominated by thermal emission, with no reflected light contributing to absorption bands near 4.38 µm or at other wavelengths. Although we think Callisto's 4.38-µm band is real, based on its prior detection in NIMS data, perhaps a subtle artifact is also contributing to the base component of its 4.38-µm band and distorting the band area ratios presented here.

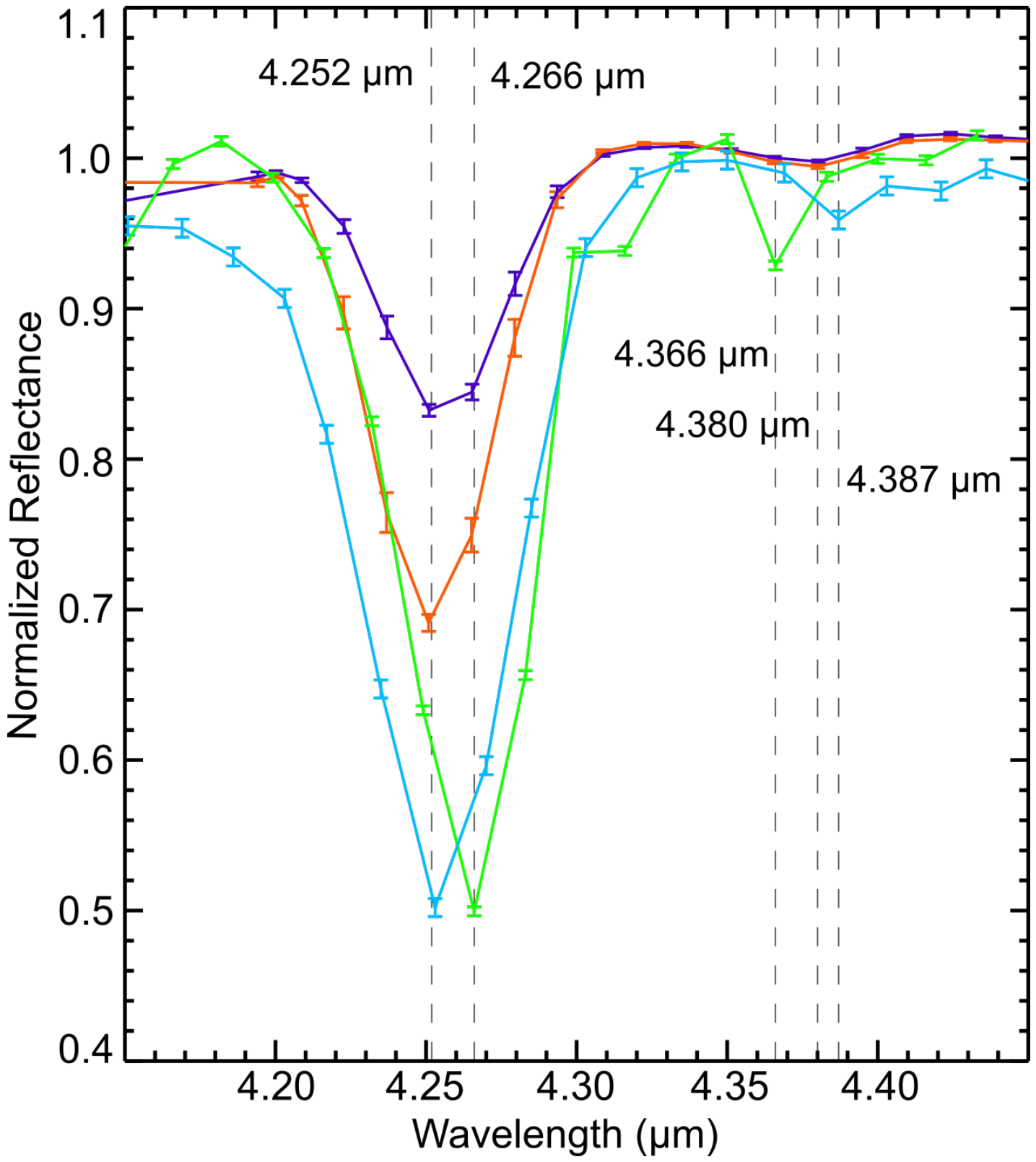

***Figure 9:*** *Comparison between $^{12}CO_2$ and $^{13}CO_2$ features in VIMS spectra of Iapetus (blue) and Phoebe (green) (Clark et al., 2019) and NIRSpec integrated spectra of Callisto's leading (purple) and trailing (orange) hemisphere, which have been binned to simulate the spectral resolution of VIMS data. All spectra are normalized to 1 at 4.35 µm. Approximate band centers are indicated (dashed lines). Error bars represent the 1σ uncertainties for each spectrum.*

To investigate this possibility, we subtracted a continuum-divided, integrated G395H spectrum of Io from the continuum-divided, integrated spectra of Callisto (hereon referred to as 'Io-subtracted' data), in the 4.38-µm wavelength region (Figure A6). We then resampled the Io-subtracted spectrum to VIMS-equivalent spectral resolutions. The resulting band area and depth ratios are notably smaller and the asymmetry between Callisto's leading and trailing hemispheres is negligible in these Io-subtracted data (<1$\sigma$ difference) (Table 3, Figure 10). Furthermore, the $^{13}CO_2/^{12}CO_2$ ratios for the Io-subtracted data are consistent with Europa's $^{13}CO_2/^{12}CO_2$ band intensity ratio (0.021 ± 0.001), measured using

NIRSpec IFU data (Villanueva et al., 2023a). We present both Io-subtracted and 'non-Io-subtracted' results in all subsequent analyses of Callisto's carbon isotope ratios.

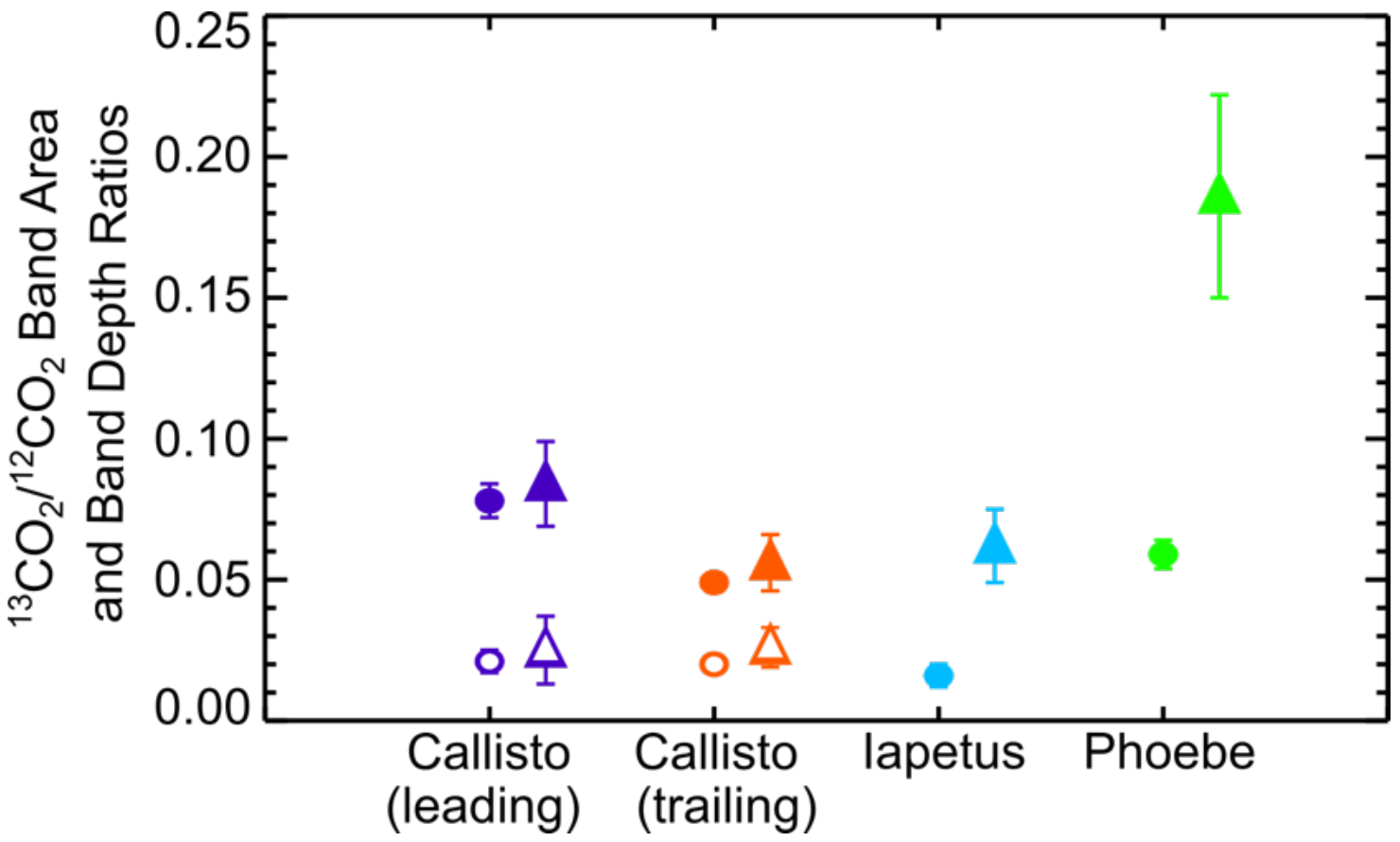

***Figure 10:*** *$^{13}CO_2/^{12}CO_2$ band area (filled circles) and depth (filled triangles) ratios and 1σ errors for the binned NIRSpec spectra of Callisto and VIMS spectra of Phoebe and Iapetus (spectra shown in Figure 9, ratios reported in Table 3). The hollow circles and triangles respectively show the band area and depth ratios for Callisto measured using Io-subtracted data.*

The band area measurements presented here essentially measure the same quantity as equivalent width, which was used in prior work showing enhanced isotopic ratios on Phoebe (Clark et al., 2019). Inverting the band area ratios reported in Table 3 (*i.e.*, $^{12}CO_2/^{13}CO_2$) provides an estimate of Callisto's $^{12}C/^{13}C$ ratio, which is a potentially useful indicator of the formation conditions in the Jovian subnebula and/or for the C-rich material delivered to Callisto's surface in dust grains. Propagating uncertainties, these $^{12}C/^{13}C$ ratios are 13 ± 1 and 20 ± 1 for Callisto's leading and trailing hemispheres, respectively, using the non-Io-subtracted spectra binned to VIMS resolution (Figure 9). In contrast, the $^{12}C/^{13}C$ ratios measured using the Io-subtracted data are 48 ± 10 and 50 ± 6 for Callisto's leading and trailing hemispheres, respectively.

Using the same technique, we calculated $^{12}C/^{13}C$ ratios of 17 ± 2 for Phoebe and 63 ± 15 for Iapetus. For context, all other solar system bodies and materials for which $^{12}C/^{13}C$ ratios has been measured in gases or solids – albeit using different techniques to those reported here – have $^{12}C/^{13}C$ ratios close to the canonical terrestrial value of 90 ± 10. These include the Sun (Lyons et al., 2018), Venus (Hoffman et al., 1980), Earth (Hauri et al., 2002), the Moon (Kerridge et al., 1975), Mars (e.g., Webster et al., 2013), Vesta (Grady et al., 1997), chondrites (e.g., Alexander et al., 2007), interplanetary dust particles (Floss et al., 2006), more than 30 comets (*e.g.*, Manfroid et al., 2009), Jupiter (Niemann et al., 1996), and Saturn (Fletcher et al., 2008). The $^{12}C/^{13}C$ ratios for Iapetus (63 ± 15) and Europa (83 ± 19; Villanueva et al. 2023a) are within this terrestrial value range, within uncertainty. A slightly larger spread is seen among comets ($^{12}C/^{13}C$ = 60 to 120, *e.g.*, Wyckoff et al., 2000), with an extreme value of 40 ± 14 for $H_2CO$ gas in comet 67P (Altwegg et al. 2020). Consequently, the $^{12}C/^{13}C$ ratios measured using the Io-subtracted data are lower than the canonical terrestrial value (~2σ difference), but fairly close to the lower range exhibited by most comets (<2σ difference) and close to Iapetus' $^{12}C/^{13}C$ ratio (<1σ difference). The $^{12}C/^{13}C$ ratios for the non-Io-subtracted Callisto data are much lower than terrestrial values (>3σ difference), as well as the ratios exhibited by most outer Solar System objects, except Phoebe.

In a scenario where $^{13}C$ is enriched on Callisto, determining whether this enrichment results from its formation conditions or its subsequent chemical evolution is challenging. Unlike Phoebe, whose high $^{13}C$ content has been ascribed to its formation history as a captured moon (Clark et al., 2019), Callisto's circular, low-inclination, prograde orbit indicates a likely formation in Jupiter's circumplanetary disk. In this case, Callisto's accreted carbon would be expected to have the

$^{12}C/^{13}C$ ratio measured for Jupiter's atmosphere, 93 ± 5 (Niemann et al. 1996), similar to $^{13}C$ on Europa (Villanueva et al., 2023a).

Formation conditions in the Jovian subnebula alone cannot explain Callisto's hemispheric dichotomy in its $^{13}CO_2/^{12}CO_2$ ratios. One possibility is that some of Callisto's dark material results from in-fall of CN-bearing debris from Jupiter's irregular satellites that are likely captured objects, possibly originating in the primordial Kuiper Belt similar to Phoebe (*e.g.*, Jewitt and Haghighipour, 2007, Nicholson et al., 2008; Nesvorný, 2018). When captured into the developing Jovian system, these satellites may have sampled out-of-midplane, $^{13}C$-rich regions of the protoplanetary disk. In such regions, the disk CO (its main carbon source) would have been dense enough for $^{12}CO$ to shield itself from photolytic UV radiation from the early Sun, thereby staying in the gas phase, whereas more tenuous $^{13}CO$ was photolyzed to products that eventually resulted in $^{13}C$-rich material (Woods & Willacy, 2009) that condensed into solids. Self-shielding of CO has similarly been invoked to explain unusual 1:1 $^{17}O/^{16}O$ and $^{18}O/^{16}O$ correlations in meteorite mineral phases (Lyons & Young, 2005), the Sun's light C isotope composition relative to the Earth (Lyons et al., 2018), and Phoebe's large enrichment in $^{13}C$ (Neveu et al., 2020). In each case, it has been assumed that the isotopically heavy material condensed into solids (ices or dust) that accreted onto, or coated, local planetesimals.

Subsequent collisions between members of the original irregular satellite population could have generated substantial amounts of debris in the form of dust grains, much of which migrated inward on decaying orbits due to Poynting-Robertson drag (Burns et al., 1979). Dust originating on retrograde irregular satellites should preferentially collide with the leading and anti-Jovian sides of the outermost regular satellite, Callisto (Bottke et al., 2010, 2013; Chen et al., 2024). Callisto's slightly darker and redder leading hemisphere (Morrison et al., 1974) has been attributed to the accumulation of red dust from the retrograde irregular satellites (Bottke et al., 2010). $^{13}C$ delivered to Callisto in dust grains should be well mixed with $H_2O$ during collisional events, representing ideal production sites for radiolytic generation of $^{13}CO_2$.

The discrepancy between our $^{13}CO_2/^{12}CO_2$ band depth and area ratios in the non-Io-subtracted data, however, highlights the need for caution when interpreting these results as Phoebe exhibits enhancement of *both* its band area and depth ratios, unlike Callisto (Figure 10). Additionally, the $^{13}CO_2/^{12}CO_2$ band ratios measured using Io-subtracted data indicate that isotopic carbon abundances on Callisto are similar to Iapetus and other bodies that exhibit terrestrial-like values, suggesting that Callisto's surface is not enriched in $^{13}C$, unlike Phoebe. Quantitative investigation of possible enhancement of $^{13}C$ via delivery of irregular satellite dust is beyond the scope of this study and left for future work. *In situ* sampling by instruments on the Europa Clipper and JUICE spacecraft of ejected dust grains and $^{13}CO_2$ molecules in Callisto's atmosphere will improve our understanding of the origin of $^{13}C$ on the Galilean satellites and refine the isotopic ratios presented here.

*4.4 Deciphering the 4.57-µm feature*

Callisto's 4.57-µm band has remained an enigma since its initial discovery by NIMS. Its broad shape and band strength points to the presence of contributing constituent(s) in fairly high abundances on Callisto's surface. Our results confirm that the 4.57-µm feature is stronger on Callisto's leading hemisphere, consistent with ground-based data (Cartwright et al., 2020). Our analysis also shows that the band center is shifted to shorter wavelengths on Callisto's trailing hemisphere compared to its leading side, suggesting that the chemical nature of the 4.57-µm feature is different on each hemisphere. At the regional scale, the 4.57-µm band is weakest near the center of Callisto's trailing side and somewhat weaker in Asgard and Valhalla compared to their surrounding terrains. Therefore, the 4.57-µm band and $^{12}CO_2$ appear to be spatially anti-associated, with more $^{12}CO_2$ present where the 4.57-µm band is weaker (Figure 11).

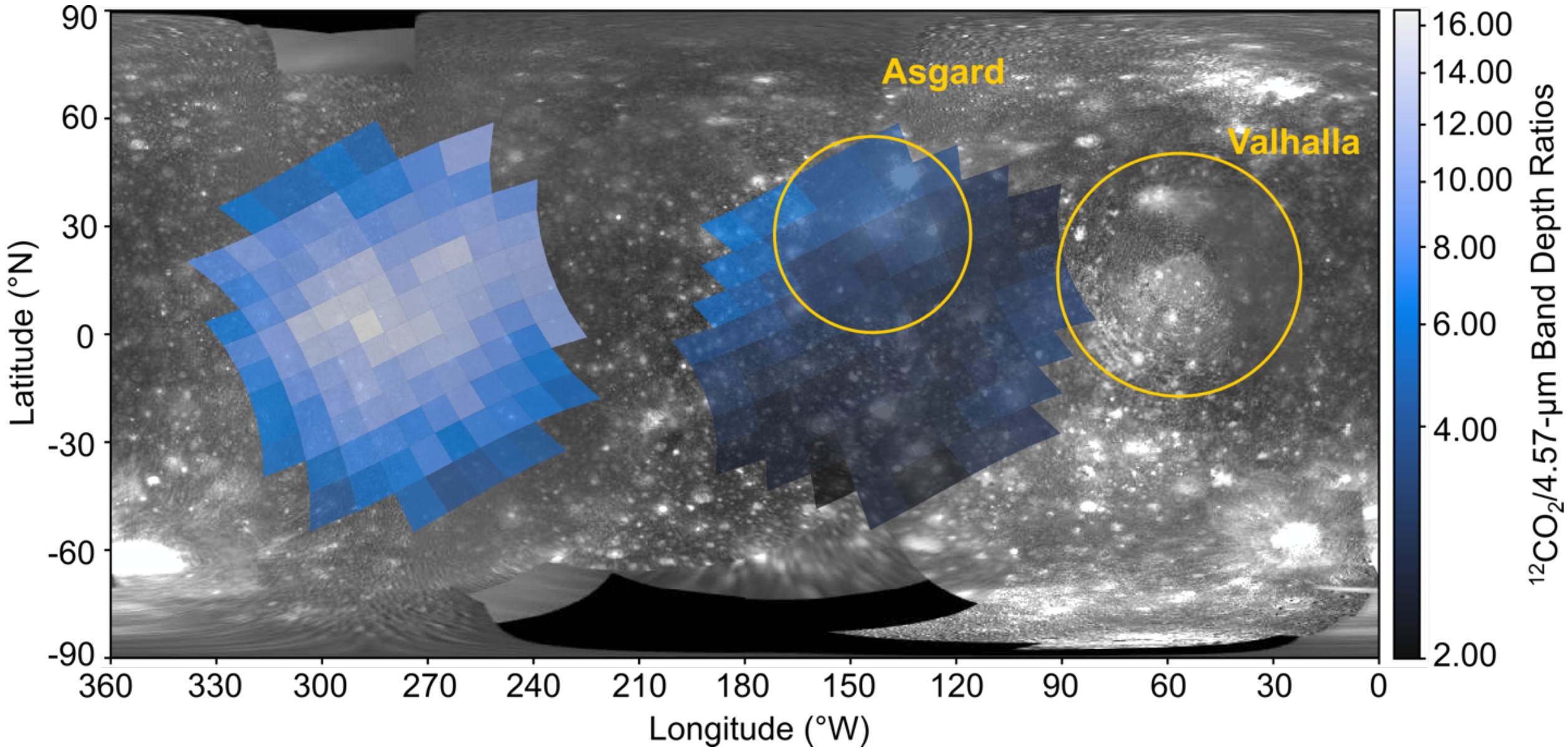

***Figure 11:*** *$^{12}CO_2$/4.57-µm band depth ratios highlighting the apparent spatial anti-association between these two spectral features. White and light blue colors indicate regions where $^{12}CO_2$ is strongest, i.e., trailing hemisphere, and dark blue colors indicate regions where the 4.57-µm band is stronger, i.e., leading hemisphere, away from Asgard and Valhalla (extents approximated with gold circles).*

This anti-association may arise from a few different scenarios. It is possible that radiolytic $CO_2$ is generated out of the C-rich components that contribute to the 4.57-µm band. In this scenario, the 4.57-µm band is likely dominated by more radiolytically-resistant components, such as CN-bearing organic residues. These CN-bearing species could have been delivered in irregular satellite dust grains (perhaps along with $^{13}C$), or delivered from other primordial objects impacting Callisto, such as micrometeorites, asteroids, and comets. Alternatively, CN-bearing organics could be native to Callisto's crust and exposed by impact events.

A 4.57-µm feature, exhibited in refractory organic residues formed via irradiation of primordial ices such as $N_2$ mixed with $CH_4$, has been measured in the laboratory and is stable at room temperature (Accolla et al., 2018). This feature does not require $H_2O$ or other O-bearing parent molecules to form it, hinting at a possibly reducing formation environment. Additionally, the 4.57-µm feature is shifted to shorter wavelengths compared to a well-documented 4.62-µm band in the interstellar medium attributed to the $OCN^-$ radical (*e.g.*, Pendleton et al., 1999; Hudson, et al., 2001; McClure et al., 2023), and these two features may be chemically distinct. Laboratory experiments continue to inform our assignment of the shorter wavelength 4.57-µm feature relative to the OCN- absorption feature and the CN fundamental stretch (Gerakines et al., 2004; Accolla et al., 2018; Gerakines et al., 2022). The presence of such organic residues on Callisto, if they formed in an $H_2O$-poor environment, supports delivery in dust grains from irregular satellites that may also be $H_2O$-poor. Upcoming JWST observations of Jovian irregular satellites (Sharkey et al., 2023a) will shed light on the possible compositional ties between these objects and Callisto's surface chemistry (Sharkey et al., 2023b), in particular for clarifying the origin of the 4.57-µm feature.

Alternatively, the 4.57-µm band could result from radiolytically generated, highly oxidized carbon molecules that form from energetic charged particles that could preferentially bombard Callisto's leading side (Liuzzo et al., 2019). In this scenario, the 4.57-µm band might result from $C_3O_2$ mixed with other carbon chain oxides (*e.g.*, $C_5O_2$, $^{13}C^{12}C_2O_2$, $C_7O_2$) that exhibit overlapping features, combining into a broad, somewhat asymmetric absorption band spanning 4.27 to 4.63 µm (*e.g.*, Trottier and Brooks, 2004; Strazzulla et al., 2007). However, irradiation of $C_3O_2$ mixed with $CO_2$, CO, or $H_2O$ shifts the center of this broad feature to between 4.46 and 4.48 µm (Gerakines and Moore, 2001), notably offset from Callisto's 4.57-µm feature. $C_3O_2$ is typically generated via irradiation of CO ices at cryogenic temperatures (~16 K; *e.g.*, Bennett et al., 2008) that are much lower than those exhibited by Callisto's surface. Although a weak CO band might exist near 4.67 µm on its trailing hemisphere (see section 4.5), this feature most likely results from ongoing radiolytic processing and is not a tracer of primordial CO ice. Furthermore, radiolytic production of $C_3O_2$ from CO would occur in conjunction with efficient production of $CO_2$, suggesting that the 4.25-µm and 4.57-µm spectral features should be spatially associated on Callisto, whereas they are apparently *anti-associated* (Figure 11). Consequently, if higher order carbon chain oxides are the primary contributors to the 4.57-µm band, then another chemical process is operating to weaken the signature of $CO_2$ where the 4.57-µm band is strongest.

Perhaps the 4.57-µm band instead results from radiolytically generated $CS_2$. It has been suggested that sulfur-bearing dust grains could be delivered from Jupiter's irregular satellites, preferentially accumulating on Callisto's leading side, mixing with C-rich material, where they are subsequently irradiated to form $CS_2$ (Cartwright et al., 2020). If S is delivered in dust and irradiated into $CS_2$, then $SO_2$ should also be present, forming from irradiation of delivered S mixed with native $H_2O$ (*e.g.*, Moore et al., 2007). Although prior analysis of NIMS data suggested a 4.05-µm band resulting from $SO_2$ is present on Callisto, as well as on Europa and Ganymede (*e.g.*, McCord et al., 1997, 1998a), subsequent analyses of ground-based datasets have found scant evidence for $SO_2$ on the icy Galilean moons (*e.g.*, Brown and Hand 2013; Cartwright et al., 2020). It has been suggested that Callisto's 4.02-µm band might result from hydrogen disulfide ($H_2S_2$), disulfanide ($HS_2$), or S allotropes ($S_2$, $S_3$, *etc*), as opposed to $SO_2$ frost that exhibits a strong feature near 4.07 µm (Cartwright et al., 2020) (summarized in Table 2). It is possible that $SO_2$ gets further processed into $SO_4$ (Carlson et al., 2002), limiting the abundance of $SO_2$ on the icy Galilean moons' surfaces. Furthermore, irradiation experiments of simple O, C, and S-bearing molecules over a range of low temperatures (10 – 150 K) found that $CS_2$ and $C_3O_2$ can form together, with $CS_2$ exhibiting a band near 6.56 µm, whereas $C_3O_2$, and not $CS_2$, dominates near 4.57 µm (Ferrante et al., 2008).

It therefore seems more likely that Callisto's 4.57-µm band is composed of native or impactor-delivered CN-bearing species and not $C_3O_2$ or $CS_2$. This finding has important implications for Callisto's chemical evolution as it supports the presence of nitrogen-bearing species, which have yet to be confirmed on Europa or Callisto and were only recently confirmed by Juno's Jupiter Infrared Auroral Mapper (JIRAM) on Ganymede in the form of $NH_4$ (Tosi et a., 2023). The CN triple bond is a possible tracer of pre-biotic chemistry and complex organics that may have been delivered to early Earth in impactors (*e.g.*, Matthews and Minard, 2006). The presence of nitriles and isonitriles on Callisto therefore could be a key tracer of habitability in the broader Jupiter system, in particular if these compounds originally accreted into Europa as it formed, or were delivered in dust and transported through Europa's icy shell into its subsurface ocean via active chaos terrains or other geologic conduits (*e.g.*, Hand et al., 2009; Hesse et al., 2022). Similar to Callisto, Phoebe exhibits a spatial anti-association between its 4.26-μm $^{12}CO_2$ feature and 4.55-μm feature, which is attributed to CN-bearing organics (Coradini et al., 2008). Future studies that compare the ~4.6-μm features on Callisto and Phoebe could reveal key insights into the nature of CN-bearing species in the outer Solar System.

If the C contributing to Callisto's 4.57-μm band is radiolytically cannibalized to form $CO_2$, the associated fate of nitrogen is not apparent. Perhaps nitrogen is eventually locked up in the form of $NH_4$, putatively detected in some ground-based observations (Calvin and Clark, 1993). Supporting evidence for $NH_4$ on Callisto in NIRSpec data or other spectral datasets is ambiguous (Table 2). As examined previously (Gerakines et al., 2022), laboratory experiments that work to elucidate the relationship between radiolytic formation and destruction of CN-bearing compounds, $CO_2$, and $OCN^-$, particularly under conditions relevant to Callisto, are needed to understand the origin and fate of the 4.57 um band and determine whether it shares chemical connections to CN-bearing organics detected elsewhere.

*4.5 Candidate constituents for other features*

*Carbon oxides?* We considered a variety of C-bearing species to explain the presence of subtle features centered near 3.28, 3.43, 3.51, 3.72, 4.67, and 4.92 μm, based on their central wavelength positions (Table 1) and visual assessment of their band shapes. A suite of laboratory experiments conducted over the past few decades have routinely demonstrated that irradiation of $H_2O$ ice mixed with carbonaceous material generates CO and $CO_2$ molecules and lower abundances of other carbon oxides, including $H_2CO_3$, $C_3O_2$, carbon trioxide ($CO_3$), dicarbon oxide ($C_2O$) and other, higher order carbon chain oxides (*e.g.*, Brucato et al., 1997; Gerakines and Moore, 2001; Mennella et al., 2004; Loeffler et al., 2005; Strazzulla et al., 2007; Ferrante et al., 2008, Bennett et al., 2010). These experiments demonstrated that many different irradiation sources (protons, electrons, heavy ions, UV photons), spanning sub-KeV to MeV energies, spur radiolytic production of carbon oxides. The $H_2O$ and C-rich surface of Callisto, orbiting within Jupiter's magnetosphere, may therefore serve as an ideal testbed for radiolytic generation of $CO_2$ and other carbon oxides.

The subtle band centered near 4.67 μm is consistent with the wavelength position of the C-O stretch of carbon monoxide. CO should be continually generated as a transitory product in a radiolytic production cycle of $CO_2$ (*e.g.*, Raut et al., 2012). The presence of CO is therefore expected on C-rich icy bodies like Callisto that are bombarded by energetic particles. However, CO ice is hyper volatile at Callisto's estimated peak surface temperatures (~170 K, Figure 5) (*e.g.*, Fray and Schmitt, 2009, and references therein). If CO is confirmed, it must be trapped in defects or pore spaces in $H_2O$ ice or hosted by some other component in Callisto's regolith, as was suggested to explain the possible presence of CO on Phoebe (Coradini et al., 2008). A similar regolith-trapping process could retain $O_2$ on Callisto and the other icy Galilean moons (*e.g.*, Spencer et al., 1995; Spencer and Calvin, 2002; Carberry Mogan et al., 2022). If CO is temporarily retained on Callisto's surface, then it might also be present in its atmosphere, albeit prior observations did not detect CO gas (Strobel et al., 2002). Future *in situ* sampling of Callisto's neutral atmosphere by MASPEX on Europa Clipper and NIM on JUICE will allow for a more sensitive search for low levels of atmospheric CO.

The broad shape of the 4.92-μm feature is consistent with the C-O symmetric stretching mode ($v_1$) in OCS measured in the laboratory and detected in the interstellar medium (*e.g.*, Palumbo et al., 1997; Ferrante et al., 2008), including recent characterization by NIRSpec in dense molecular clouds (McClure et al., 2023). OCS also displays an absorption band near 3.4 µm that might explain Callisto's 3.43-µm band, as suggested by a prior ground-based study (Cartwright et al., 2020). The 4.92-µm feature may also result from $CO_3$, which forms from irradiated CO and $CO_2$. $CO_3$ can form alongside OCS in substrates composed of carbon oxides and sulfur-bearing species, with both species contributing to 4.9-µm features that are difficult to untangle (see Fig. 5 in Ferrante et al., 2008). Alternatively, crystalline $^{12}CO_2$ ice can exhibit a 4.9-µm band (*e.g.*, Hansen, 1997), but it is uncertain whether this feature would be expressed by the complexed $CO_2$ that dominates Callisto's surface.

*Na-bearing species?* Several spectral features on Callisto hint at the possible presence of Na-bearing minerals. The broad 'elbow' shaped feature centered near 3.72 µm corresponds to a modest change in the slope of Callisto's continuum slope between ~3.7 to 3.8 µm. Oxalates ($C_2O_4$-bearing species) like natroxalate ($Na_2C_2O_4$) (Applin et al., 2016) and sulfates ($SO_4$-bearing salts) such as thenardite ($Na_2SO_4$) (De Angelis et al., 2021) can exhibit broad features in the 3.6 to 3.8 µm wavelength range. Similar to Ceres' strong 4-µm band (*e.g.*, De Sanctis et al., 2016; Carrozzo et al., 2018; Raponi et al., 2019), Callisto's prominent 4.02-µm band has been attributed to Na-bearing carbonates (Johnson et al., 2004). Additionally, Callisto's subtle 3.43-µm and 3.51-µm features could be weak tracers of $Na_2CO_3$ or other carbonates. Na-bearing species have been implicated on the surface of Europa in the form of Na carbonates (*e.g.*, McCord et al., 1998b) and irradiated NaCl (*e.g.*, Trumbo et al., 2019, 2022) that are predicted to have originated in its internal ocean (*e.g.*, Hand and Carlson, 2015) and from Na delivered to Europa via volcanic outgassing on Io (*e.g.*, Carlson et al., 2009). By extension, perhaps Na-bearing components are native to Callisto, accreting into its interior as it formed in the Jovian subnebula. In this scenario, Na-rich material in the crust might form salts, if in contact with, and subsequently extracted from, pockets of liquid water in Callisto's early history. Alternatively, perhaps Na salts could form in response to aqueous alteration driven by impact-induced melting (Yasui et al., 2021), assuming Na is available in Callisto's crust or delivered in impactors. Once formed, Na salts could be subsequently exposed by impact gardening.

*Organics:* The subtle absorption bands near 3.28, 3.43, and 3.51 µm might result from short-chain CH-bearing organics (*i.e.*, hydrocarbons), as suggested in prior work (McCord et al., 1997, 1998a). Radiolysis of surface hydrocarbons has been implicated as a possible source of the $H_2$ detected in Callisto's atmosphere (Carberry Mogan et al., 2022). These weak features show comparable band strengths on Callisto's leading and trailing side ($<1\sigma$ difference), suggesting the constituents that contribute to them may be well mixed in Callisto's dark regolith, perhaps serving as source material for radiolytic $CO_2$. Such CH-bearing components of functional groups may also be incorporated into CN-bearing constituents in long-chain refractory organic residues (complex organic molecules similar to laboratory tholins) that have been implicated for the 4.57-µm band (McCord et al., 1998a). Additionally, a feature detected near 3.65 µm in Jupiter Infrared Auroral Mapper data of Ganymede has been attributed to aldehydes (Tosi et al., 2023), and perhaps these species are contributing to Callisto's weak 3.72-µm feature (albeit there is a large wavelength difference).

*Future work:* Subsequent studies that more rigorously compare observed spectral features to laboratory spectra of a wide range of C and S-bearing constituents are needed to better understand Callisto's surface chemistry. These six features are fairly weak and confirmation of their presence with JWST and other telescopes is needed. Furthermore, high spatial resolution mapping of Callisto's spectral properties during upcoming close flybys by NASA's Europa Clipper and ESA's

JUICE spacecraft will be key to understanding its surface chemistry. Such spectral maps could be used to identify possible spatial associations between different spectral features. For example, a spatial association between the broad 4.57-µm band and subtle features centered near 3.28, 3.43, and 3.51 µm would suggest these features result from refractory organic residues that exhibit C-H and C-N stretching modes. Alternatively, spatial associations between the 3.43-µm and 3.51-µm bands, along with the 3.72-µm, 4.02-µm, and 4.92-µm bands, would suggest they result from carbonates/$CO_3$. If only the 3.43-µm and 4.92-µm bands are associated, they could result from OCS.

*4.6 A $CO_2$ cycle on Callisto*

The results reported here confirm that Callisto is a world dominated by carbon and $H_2O$, with minor amounts of S-bearing species likely present as well. Landscape evolution models suggest the degraded nature of craters on Callisto, formation of icy pinnacles, and the numerous examples of mass wasting features, result from the sublimation of $H_2O$ ice and crustal $CO_2$ ice exposed at the surface or retained in its near-surface (White et al., 2016). In this scenario, solid-state $CO_2$ should be gradually transferred from Callisto's surface and subsurface to its atmosphere. Most of this atmospheric $CO_2$ should eventually condense and get trapped in Callisto's regolith, with only a small fraction sufficiently accelerated by interactions with magnetospheric ions to experience Jeans escape. The presence of $CO_2$ gas across Callisto's disk supports this surface-atmosphere transfer process, which may be enhanced in some regions, such as Valhalla and a large plains unit on its trailing side (Figure 5, section 4.2). A similar process might be occurring for atmospheric $O_2$ which is potentially enriched by transfer of $O_2$ molecules weakly bound in defects and on grain surfaces in Callisto's porous regolith (Carberry Mogan et al., 2022).

Some fraction of atmospheric $CO_2$ should migrate and condense on colder landforms, in particular on Callisto's nightside, hypothetically contributing to the growth of icy pinnacles (or at least temporarily cold trapping on them). Unlike Europa or Ganymede, there is little evidence for widespread endogenic activity on Callisto, and its surface is likely only geologically refreshed by impacts. Consequently, Callisto has built up a thick lag deposit of dark material, which might be isotopically heavy, as well as rich in amorphous C and other possible radiolytic end products, formed by continual charged particle bombardment of organics. New impact events can puncture this dark blanket, mixing crustal $H_2O$ ice and other components with ancient, irradiated regolith material, perhaps providing new radiolytic production sites for $CO_2$ molecules. This process could help replenish Callisto's inventory of solid-state $CO_2$. Such a process could also expose Na-bearing salts retained in the crust that might react with $S^{n+}$ ions to form sulfates and other S-bearing species, as suggested for the proposed exogenic formation of Mg-sulfates on Europa (Brown and Hand, 2013).

A key test of $CO_2$ surface-atmosphere cycling is whether crystalline $CO_2$ ice is present and spatially associated with impact features and other landforms that show exposed crustal materials. Although we do not directly detect 'pure' $CO_2$ ice, the shifted central wavelength of the complexed $CO_2$ band in spaxels associated with Asgard and Valhalla could result from the presence of minor amounts of crystalline $CO_2$. Alternatively, another component mixed with $CO_2$, such as amorphous $H_2O$ ice (Bockelée-Morvan et al., 2023) or perhaps the CN-bearing organics that may contribute to the 4.57-µm band, might explain the wavelength shift of Callisto's $^{12}CO_2$ band. These contaminants could also contribute to the broad base component of Callisto's $^{13}CO_2$ band. Laboratory experiments conducted under conditions relevant to Callisto are needed to substantiate these possibilities. Furthermore, close passes made by spacecraft with near-infrared spectrometers could look for the spectral signature of crystalline $CO_2$ ice, in particular in fresh craters where exposed crustal deposits might still be present. Albeit even if a clear association between $CO_2$ ice and fresh craters is eventually established in spacecraft datasets, additional tests will be needed to discern between native $CO_2$ ice and condensed $CO_2$ ice that is formed elsewhere and subsequently cold trapped on fresh craters that tend to be brighter and colder.

### *4.7 Comparing Callisto to other icy bodies*

*$CO_2$ on Callisto, Ganymede, and Europa:* The icy Galilean moons all exhibit spectral features consistent with the $v_3$ mode of $^{12}CO_2$, which were originally detected by Galileo/NIMS (Carlson et al., 1996) and confirmed by JWST/NIRSpec observations in 2022 (Figure 12). When comparing the integrated NIRSpec spectra of these moons, Callisto displays the strongest $^{12}CO_2$ band, with band depths ranging between 15 to 40% of the continuum, Ganymede's $^{12}CO_2$ feature exhibits band depths between 9 to 19% of the continuum (Bockelée-Morvan et al., 2023), and Europa's $^{12}CO_2$ feature has band depths between 5 to 10% (Villanueva et al., 2023a). Callisto's trailing hemisphere and Europa both display 4.25-µm $CO_2$ bands that are remarkably similar in central position (Figure 12), whereas Ganymede's $^{12}CO_2$ band is shifted closer to 4.26 µm. Europa also displays a 4.27-µm band consistent with crystalline $CO_2$ ice, which is absent from Callisto, but Ganymede's $^{12}CO_2$ band center does shift to ~4.27 µm at high north polar latitudes, possibly because $CO_2$ molecules are trapped in amorphous $H_2O$ ice (Bockelée-Morvan et al., 2023). Similarly, the shifted position of Ganymede's and Callisto's $^{12}CO_2$ band at low latitudes on their leading sides might result from $CO_2$ trapped in $H_2O$ ice (Bockelée-Morvan et al., 2023), or alternatively, perhaps minor amounts of crystalline $CO_2$ ice are mixed in with the stronger complexed $CO_2$ band, thereby convolving the two features.

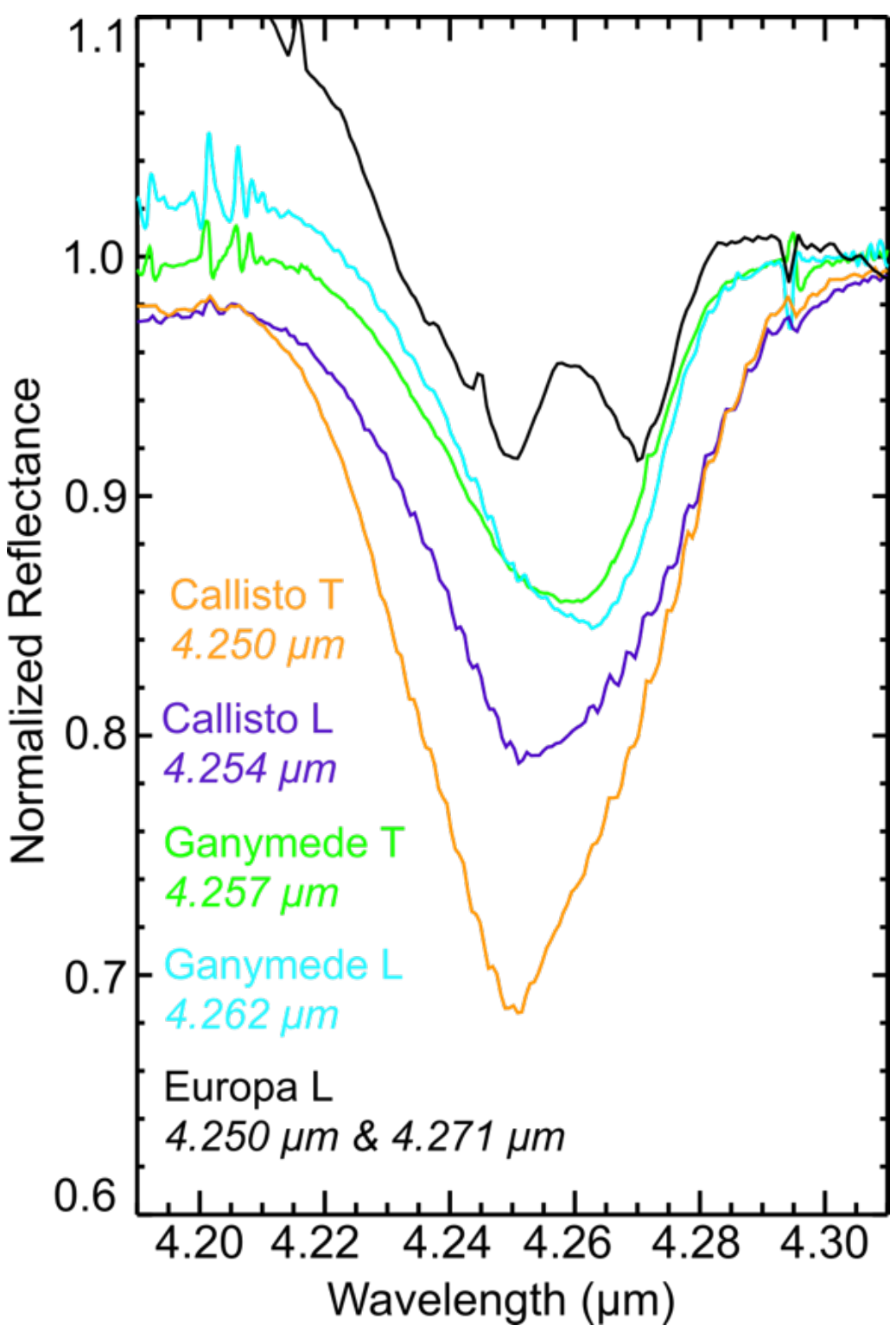

***Figure 12:*** *Integrated NIRSpec spectra on Callisto's leading (purple) and trailing (orange), Ganymede's leading (blue) and trailing (green), and Europa's leading (black) hemisphere highlighting the $^{12}CO_2$ band, normalized to 1 at 4.305 µm. Central wavelengths for each $^{12}CO_2$ band are listed on the plot using the same color scheme. Error bars have been suppressed for clarity.*

Although NIRSpec confirmed the presence of $CO_2$ gas rovibrational lines at Callisto, measured earlier by NIMS (Carlson 1999), these emission features were not detected by NIRSpec at Europa (Villanueva et al. 2023a). The reason for the apparent disparity in the presence of $CO_2$ gas is uncertain, but the higher peak surface temperatures and larger surface inventories of $CO_2$ on Callisto could help maintain an ongoing $CO_2$ sublimation-condensation cycle that is more difficult to sustain at Europa and perhaps Ganymede, limiting the amount of $CO_2$ gas in their atmospheres. Albeit peak $CO_2$ gas abundances do not appear to be associated with the subsolar region on Callisto (Figure 5), highlighting the complexities in the processes that form and sustain its $CO_2$ atmosphere. Additionally, differences in the geochemical evolution of volatile components in the near-surface layers of these moons, due to differences in geologic activity and associated surface ages, may play a role as well.

*Are Callisto and Ceres spectral analogs?* Although the main belt asteroid Ceres is smaller than Callisto, experiences higher peak temperatures (~235 K at the subsolar point; Tosi et al., 2015), and does not orbit within the massive Jovian magnetosphere, they appear to share some interesting compositional similarities. Ceres has a surface rich in carbonates, such as $Na_2CO_3$ (*e.g.*, De Sanctis et al., 2016; Carrozzo et al., 2018; Raponi et al., 2019), and hydrated minerals like $NH_4$-bearing phyllosilicates (*e.g.*, Ammannito and Ehlmann, 2022), which might be present on Callisto. Both worlds have ancient, heavily cratered surfaces with deposits of more volatile components in impact features, including $H_2O$ ice on Ceres (Platz et al., 2016; Combe et al., 2016, 2019) and solid-state $CO_2$ on Callisto. Additionally, dark material on Ceres likely includes organics (De Sanctis et al., 2017; Prettyman et al., 2017; Marchi et al., 2019). One possible reason why Ceres and Callisto exhibit similar spectral properties is that Ceres may have formed in the outer Solar System and was subsequently scattered into the main belt (e.g., Raymond and Izidoro, 2017; Raymond and Nesvorný, 2020; de Sousa et al., 2022). In this scenario, Ceres and Callisto might have accreted similar inventories of starting components, with Callisto eventually retaining more $H_2O$ due to its colder environment. Further consideration of how these two worlds have chemically evolved would improve our understanding of the spectrum of ocean worlds, with less active Callisto and Ceres perhaps representing a 'geologic bookend' to hyperactive Enceladus and Europa. Future close flybys of Callisto by the JUICE and Europa Clipper spacecraft will enable better spectral and geologic comparison to Ceres and other bodies at far higher spatial resolutions than can be achieved by JWST.

## 5. Conclusions

We analyzed integrated spectra and band maps of Callisto, made using JWST/NIRSpec IFU spectral cubes (G395H). These high S/N data confirmed that Callisto's surface has a large inventory of solid-state $^{12}CO_2$, indicated by a prominent 4.25-µm absorption band that was previously attributed to complexed $CO_2$, where this component is molecularly bound to more refractory species, allowing it to persist at Callisto's peak surface temperatures (~170 K). This broad $CO_2$ band, spanning 4.2 to 4.3 µm, is overprinted by $CO_2$ gas rovibrational emission lines in the NIRSpec data, confirming the previously estimated abundance of $CO_2$ in Callisto's atmosphere. Our $CO_2$ gas density estimates are a factor of 2 to 10 lower than the estimated density of the (likely) main atmospheric constituent $O_2$. The signature of solid-state $CO_2$ is significantly stronger on Callisto's trailing hemisphere, peaking at low latitudes near its apex, and steadily decreasing from this central location. On Callisto's leading hemisphere, the signature of $CO_2$ is weaker, peaking in locations associated with the Asgard and Valhalla impact basins, suggesting it is influenced by regional geologic terrains. The central wavelength of the $CO_2$ band is also distinct on each hemisphere, centered close to 4.25 µm on Callisto's trailing side but shifted to 4.258 µm in spaxels covering Asgard and Valhalla.

The 'bullseye' pattern distribution on its trailing side is consistent with radiolytic production of $CO_2$ molecules, out of native carbonaceous species mixed with $H_2O$, via interactions with Jupiter's corotating plasma that preferentially flows onto the trailing hemispheres of the Galilean moons. The overall weaker $^{12}CO_2$ features and more mottled distribution on its leading side is more consistent with exposure of crustal $CO_2$ in impact features and/or cold trapping of $CO_2$ sourced from elsewhere. An ongoing sublimation-condensation cycle could be operating on Callisto, where $CO_2$ diffuses out of its dark regolith, gets transported in its tenuous atmosphere, and condenses on bright and relatively cold crater rims, ejecta blankets, and icy pinnacles (at least temporarily). Of note, peak $CO_2$ gas abundances do not coincide with the subsolar region on either hemisphere, highlighting that sputtering and radiolytic processes and interactions with large geologic terrains likely help sustain Callisto's $CO_2$ atmosphere.

Analysis of the NIRSpec data confirmed the presence of a 4.38-µm band that likely results from $^{13}CO_2$. This feature may be spuriously broadened by a calibration artifact. However, Callisto's 4.38-µm band is notably stronger in some spaxels (~5% band depths) than the calibration artifacts possibly detected on other targets (≲2% band depths). We measured Callisto's $^{13}CO_2$/$^{12}CO_2$ band area and depth ratios and compared them to the Saturnian moons Phoebe, which is enriched in $^{13}C$, and Iapetus, which exhibits terrestrial-like values of $^{13}C$ (Clark et al., 2019). We found that Callisto's $^{13}CO_2$/$^{12}CO_2$ band area ratios are similar to Phoebe, suggesting that Callisto's surface may be enriched in $^{13}C$. In contrast, Callisto's $^{13}CO_2$/$^{12}CO_2$ band depth ratios are consistent with Iapetus, suggesting a terrestrial-like abundance of $^{13}C$ on Callisto. The possible broadening of this feature by a calibration artifact may spuriously enhance its band area, suggesting that the band depth ratios could be more reliable. The disparity between these band ratios requires additional work to better constrain Callisto's surface inventory of $^{13}C$.

We confirmed the presence of a broad 4.57-µm absorption band that is significantly stronger on Callisto's leading hemisphere and appears to be anti-associated with the distribution of solid-state $^{12}CO_2$. This feature could result from CN-bearing organics that are native to Callisto and/or delivered in dust grains from the irregular satellites, which may serve as source material that is consumed by the radiolytic production cycle forming complexed $CO_2$. We identified five other absorption features detected previously, centered near 3.28, 3.43, 3.51, 3.72, and 4.02 µm, which could result from C-bearing species, such as organics and carbonates. We also detected two other absorption features for the first time on Callisto, centered near 4.67 µm and 4.92 µm, possibly resulting from CO and OCS, respectively. These JWST/NIRSpec observations reinforce the existing body of work that indicates Callisto's surface exhibits complex geological and chemical processing of C-rich material, evidenced by the ubiquitous presence of $CO_2$ that is likely cycling between its surface and atmosphere. Follow up observations by JWST/NIRSpec and other telescope facilities and instruments are needed to corroborate these subtle features.

**Table 1:** Band measurements for the integrated spectra

| Band Name | Band Center (µm) | Band Wavelength Range (µm) | Hemisphere | Band Depth (%) | Band Area ($10^{-4}$ µm) | >3σ Band Depth and Area? |
|---|---|---|---|---|---|---|
| 3.28-µm | 3.278 | 3.223 - 3.319 | Leading | 1.27 ± 0.29 | 4.95 ± 0.56 | Yes |
| | | | Trailing | 1.32 ± 0.18 | 4.30 ± 0.31 | Yes |
| 3.43-µm | 3.428 | 3.410 - 3.445 | Leading | 1.05 ± 0.24 | 1.63 ± 0.27 | Yes |
| | | | Trailing | 0.97 ± 0.20 | 1.28 ± 0.19 | Yes |
| 3.51-µm | 3.512 | 3.494 - 3.529 | Leading | 0.63 ± 0.19 | 0.91 ± 0.24 | Yes |
| | | | Trailing | 0.90 ± 0.16 | 0.92 ± 0.18 | Yes |
| 3.72-µm | 3.720 | 3.686 - 3.769 | Leading | 0.07 ± 0.20 | 0.039 ± 0.35 | No |
| | | | Trailing | 0.76 ± 0.17 | 3.10 ± 0.28 | Yes |
| *4.02-µm | 4.012 | 3.950 - 4.050 | Leading | 2.67 ± 0.15 | 12.80 ± 0.28 | Yes |
| | | | Trailing | 0.64 ± 0.27 | 1.26 ± 0.26 | No |
| 4.25-µm | $^{L}$4.255 | 4.208 - 4.305 | Leading | 18.90 ± 0.16 | 83.54 ± 0.35 | Yes |
| | $^{T}$4.250 | | Trailing | 33.23 ± 0.46 | 148.92 ± 0.21 | Yes |
| *†4.30-µm* | *4.295* | *4.293 - 4.302* | *Leading* | *1.16 ± 0.45* | *0.19 ± 0.11* | *No* |
| | | | *Trailing* | *1.48 ± 0.29* | *0.34 ± 0.09* | *Yes* |
| 4.38-µm | 4.382 | 4.335 - 4.415 | Leading | 2.31 ± 0.27 | 7.54 ± 0.34 | Yes |
| | | | Trailing | 2.86 ± 0.20 | 8.76 ± 0.27 | Yes |
| 4.57-µm | 4.565 | 4.487 - 4.619 | Leading | 5.98 ± 0.21 | 40.53 ± 0.42 | Yes |
| | | | Trailing | 2.78 ± 0.19 | 20.87 ± 0.35 | Yes |
| 4.67-µm | 4.670 | 4.648 - 4.699 | Leading | 0.35 ± 0.20 | 0.89 ± 0.28 | No |
| | | | Trailing | 0.71 ± 0.16 | 1.70 ± 0.21 | Yes |
| 4.92-µm | 4.922 | 4.835 - 4.969 | Leading | 0.87 ± 0.21 | 7.27 ± 0.48 | Yes |
| | | | Trailing | 1.61 ± 0.21 | 11.70 ± 0.37 | Yes |
| *††5.00-µm* | *4.996* | *4.975 - 5.010* | *Leading* | *2.49 ± 0.19* | *2.32 ± 0.25* | *Yes* |
| | | | *Trailing* | *1.91 ± 0.26* | *1.61 ± 0.21* | *Yes* |
| *††5.04-µm* | *5.042* | *5.025 - 5.056* | *Leading* | *1.01 ± 0.22* | *1.44 ± 0.23* | *Yes* |
| | | | *Trailing* | *0.91 ± 0.23* | *1.34 ± 0.18* | *Yes* |
| *††5.07-µm* | *5.074* | *5.060 - 5.084* | *Leading* | *1.12 ± 0.24* | *1.45 ± 0.21* | *Yes* |
| | | | *Trailing* | *1.06 ± 0.21* | *1.39 ± 0.16* | *Yes* |

*$^{L}$ = Leading, $^{T}$ = Trailing*

**Band measured using smaller subset of 18 spaxels (leading) and 15 spaxels (trailing).*

*†Possible residual solar line.*

*††Possible artifact.*

**Table 2:** Absorption bands detected between 3 and 5 μm and possible constituents.

| Band Name | Detected in Integrated NIRSpec Data? | Detected in NIMS Data? | Detected in Ground-based Data? | Constituents (confirmed, bolded) (suggested, italicized) | References |
|---|---|---|---|---|---|
| 3.00-μm | Yes | Yes | Yes | **$H_2O$**<br>(ice & hydrates) | Pollack et al. (1978) |
| 3.05-μm | †No | Yes | Yes | *$NH_4$-bearing*<br>*OH-bearing* | Calvin and Clark (1993)<br>Moore et al. (2004) |
| 3.10-μm | Yes | Yes | Yes | **$H_2O$ ice** | Calvin and Clark (1993) |
| 3.28-μm | Yes | Yes | No | *CH-bearing*<br>*Carbonates* | Moore et al. (2004)<br>This Work |
| 3.43-μm | Yes | Yes | Yes | *$H_2O$ frost*<br>*CH-bearing*<br>*Carbonates*<br>*OCS* | Calvin and Clark (1993)<br>McCord et al. (1998a)<br>*McCord et al. (1998a)<br>Cartwright et al. (2020a) |
| 3.51-μm | Yes | Yes | Yes | *CH-bearing*<br>*Carbonates* | McCord et al. (1998a)<br>*McCord et al. (1998a) |
| 3.72-μm | Yes | Yes | Yes | *Oxalates*<br>*Sulfates* | This Work<br>Cartwright et al. (2020a) |
| 3.88-μm | †No | Yes | ††Yes | *$H_2CO_3$*<br>*$H_2S$* | Johnson et al. (2004)<br>McCord et al. (1997, 1998a) |
| 4.02-μm | Yes | Yes | Yes | *$SO_2$*<br>*Carbonates*<br>*S-allotropes, $H_2S_2$, $HS_2$* | McCord et al. (1997, 1998a)<br>Johnson et al. (2004)<br>Cartwright et al. (2020a) |
| 4.12-μm | †††No | Yes | No | *$D_2O$, HDO* | Clark et al. (2019) |
| 4.25-μm | Yes | Yes | No | **$^{12}CO_2$** | Carlson et al. (1996) |
| 4.30-μm | Yes | No | No | *$^{16}O^{12}C^{18}O$*<br>*Residual solar line* | This Work |
| 4.38-μm | Yes | Yes | No | **$^{13}CO_2$** | McCord et al. (1998a) |
| 4.57-μm | Yes | Yes | Yes | *CN-bearing*<br>*$C_3O_2$*<br>*$CS_2$* | McCord et al. (1997, 1998a)<br>Johnson et al. (2004)<br>Cartwright et al. (2020a) |
| 4.67-μm | Yes | No | No | *CO* | This Work |
| 4.92-μm | Yes | No | No | *OCS*<br>*$CO_3$*<br>*$^{12}CO_2$* | This Work |

*†Feature may be present in some individual NIRSpec spaxels.*

*††Feature is much weaker and narrower in ground-based data compared to NIMS data.*

*†††Feature is entirely within the unrecoverable wavelength gap of the G395H.*

**Carbonates were considered but ruled out based on the available spectral libraries.*

**Table 3:** $CO_2$ band measurement ratios

| Observation | $^{12}CO_2$ Central Wave. (µm) | $^{12}CO_2$ Band Depth (%) | $^{12}CO_2$ Band Area ($10^{-4}$ µm) | $^{13}CO_2$ Central Wave. (µm) | $^{13}CO_2$ Band Depth (%) | $^{13}CO_2$ Band Area ($10^{-4}$ µm) | $^{13}CO_2/^{12}CO_2$ Band Depth Ratio | $^{13}CO_2/^{12}CO_2$ Band Area Ratio |
|---|---|---|---|---|---|---|---|---|
| **NIRSpec Data** | | | | | | | | |
| Callisto, Leading | 4.254 | 18.90 ± 0.16 | 83.54 ± 0.35 | 4.382 | 2.31 ± 0.27 | 7.54 ± 0.34 | 0.122 ± 0.014 | 0.090 ± 0.004 |
| | | | | | †0.54 ± 0.11 | †2.85 ± 0.38 | †0.029 ± 0.006 | †0.034 ± 0.005 |
| Callisto, Trailing | 4.250 | 33.23 ± 0.46 | 148.92 ± 0.21 | 4.382 | 2.86 ± 0.20 | 8.76 ± 0.27 | 0.086 ± 0.006 | 0.059 ± 0.002 |
| | | | | | †0.81 ± 0.11 | †3.70 ± 0.37 | †0.024 ± 0.003 | †0.025 ± 0.003 |
| **VIMS Data** | | | | | | | | |
| Phoebe | 4.266 | 42.55 ± 7.79 | 270.07 ± 2.57 | 4.366 | 7.91 ± 0.55 | 15.83 ± 1.47 | 0.186 ± 0.036 | 0.059 ± 0.005 |
| Iapetus | 4.254 | 43.78 ± 4.74 | 289.53 ± 2.59 | 4.387 | 2.71 ± 0.47 | 4.60 ± 1.06 | 0.062 ± 0.013 | 0.016 ± 0.004 |
| *Callisto, Leading | 4.252 | 14.56 ± 0.74 | 68.15 ± 1.97 | 4.380 | 1.23 ± 0.21 | 5.34 ± 0.36 | 0.084 ± 0.015 | 0.078 ± 0.006 |
| | | | | | †0.37 ± 0.18 | †1.41 ± 0.30 | †0.025 ± 0.012 | †0.021 ± 0.004 |
| *Callisto, Trailing | 4.252 | 25.83 ± 3.23 | 128.38 ± 3.63 | 4.380 | 1.46 ± 0.19 | 6.31 ± 0.38 | 0.056 ± 0.010 | 0.049 ± 0.003 |
| | | | | | †0.67 ± 0.16 | †2.55 ± 0.30 | †0.026 ± 0.007 | †0.020 ± 0.002 |

*†Io spectrum subtracted from integrated Callisto spectra (Figure A6).*

**Integrated Callisto spectra resampled to simulate VIMS resolving power at 4.3 µm (Figure 9).*

## Appendix

### *A.1 Band depth and center error maps*

Here we report the error maps for band depth and center distribution plots for the 4.25-µm $^{12}CO_2$ band, 4.38-µm $^{13}CO_2$ band, and 4.57-µm band shown in Figures 6, 7, and 8, respectively.

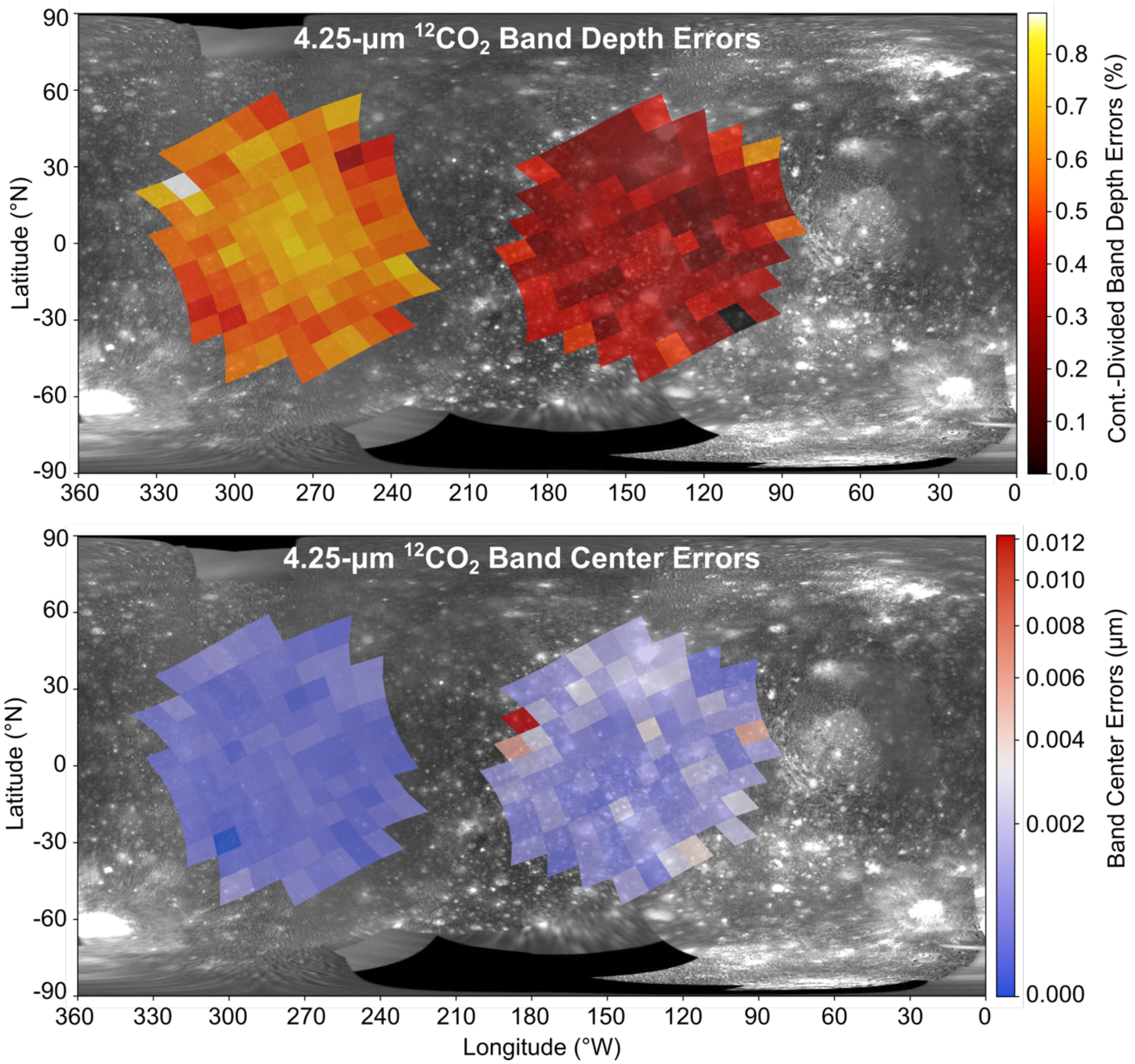

***Figure A1:*** *4.25-µm $^{12}CO_2$ band depth (top) and band center (middle) error maps for data shown in Figure 6.*

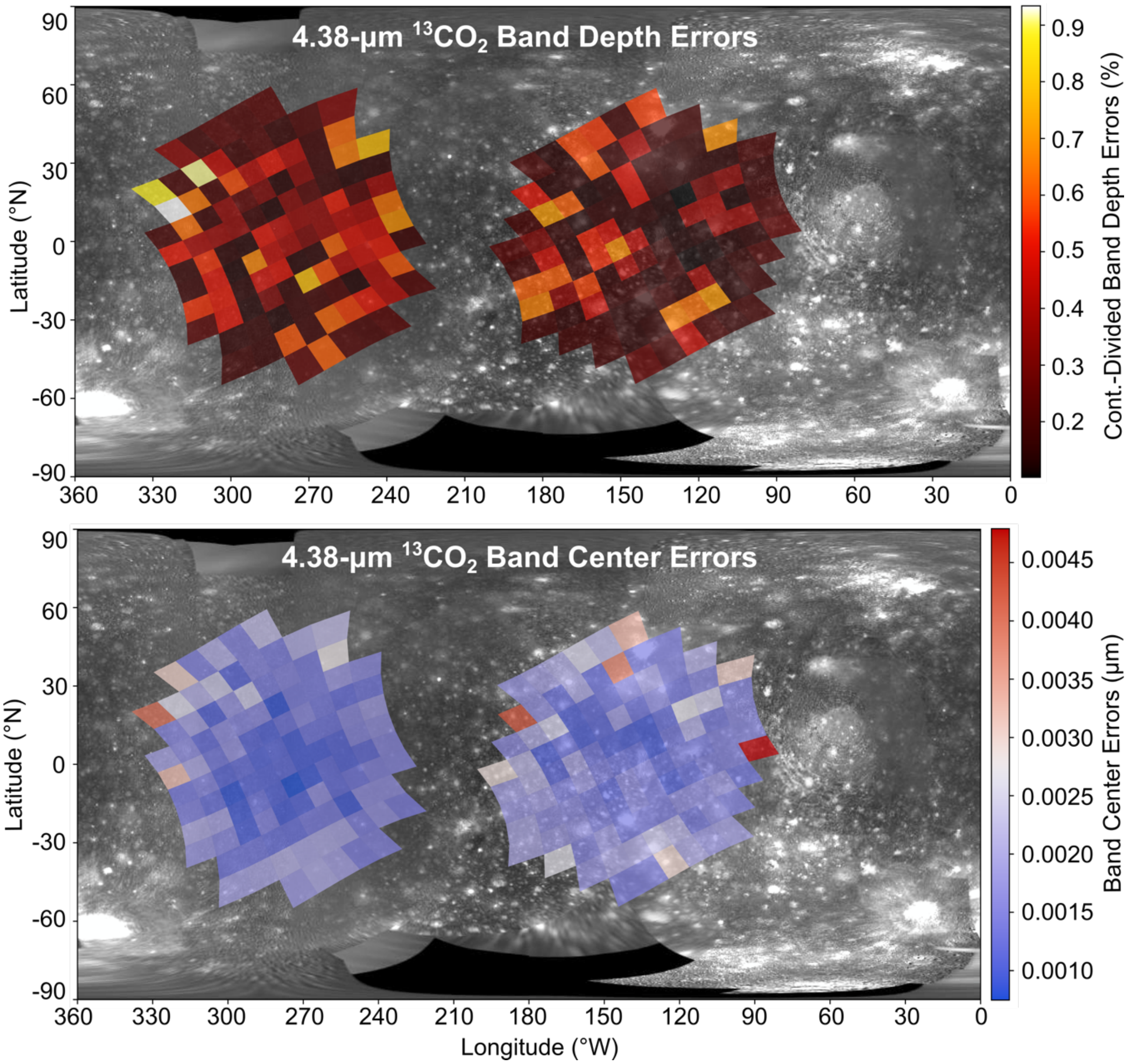

***Figure A2:*** *4.38-μm $^{13}CO_2$ band depth (top) and band center (bottom) error maps for data shown in Figure 7.*

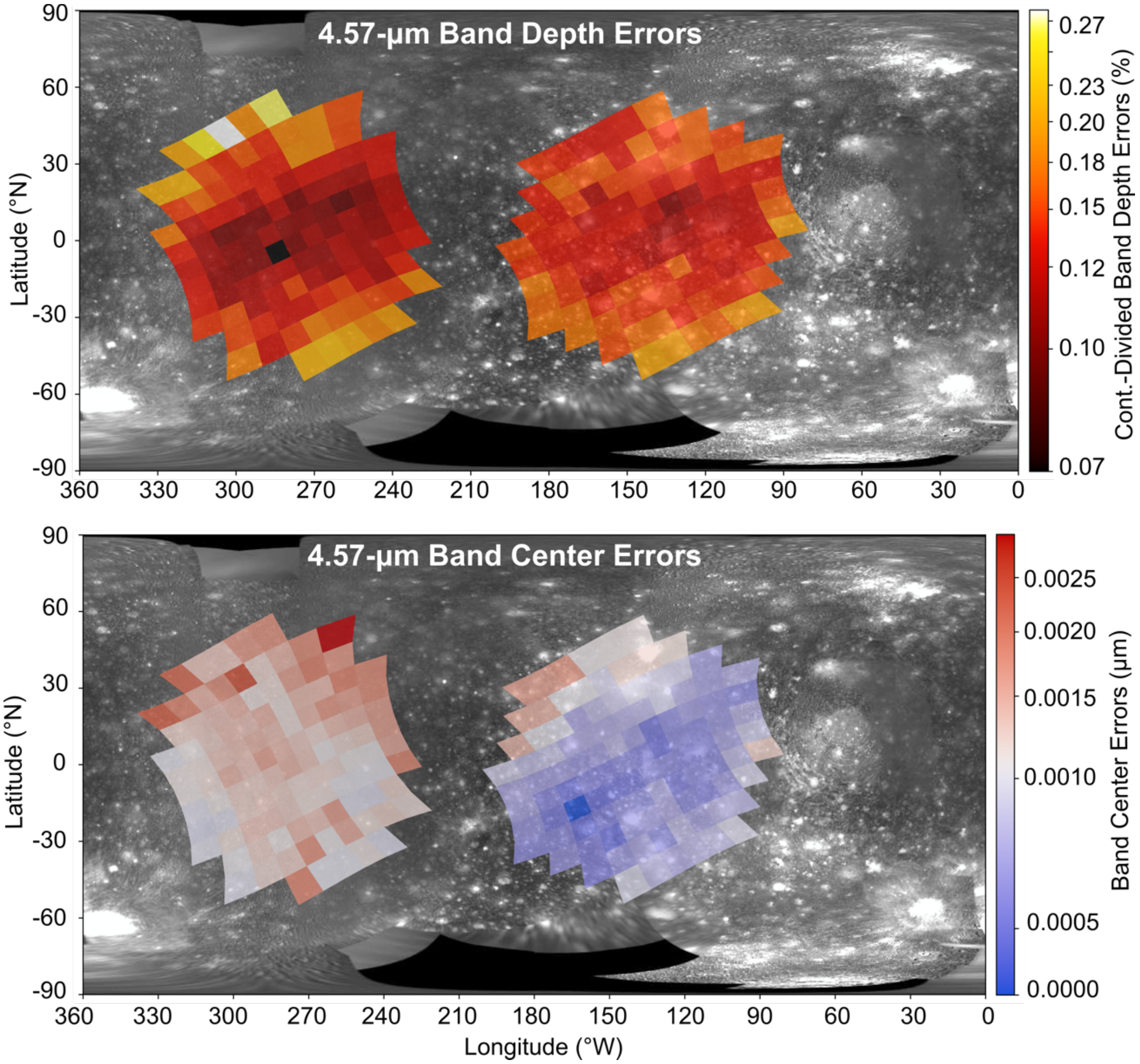

***Figure A3:*** *4.57-µm band depth (top) and band center (bottom) error maps for data shown in Figure 8.*

*A.2 Band depth map for the 4.38-µm feature*

Here we report the band depth map for the 'peak' component of the 4.38-µm $^{13}CO_2$ band, complementing the base + peak components of the 4.38-µm $^{13}CO_2$ band shown in Figure 7.

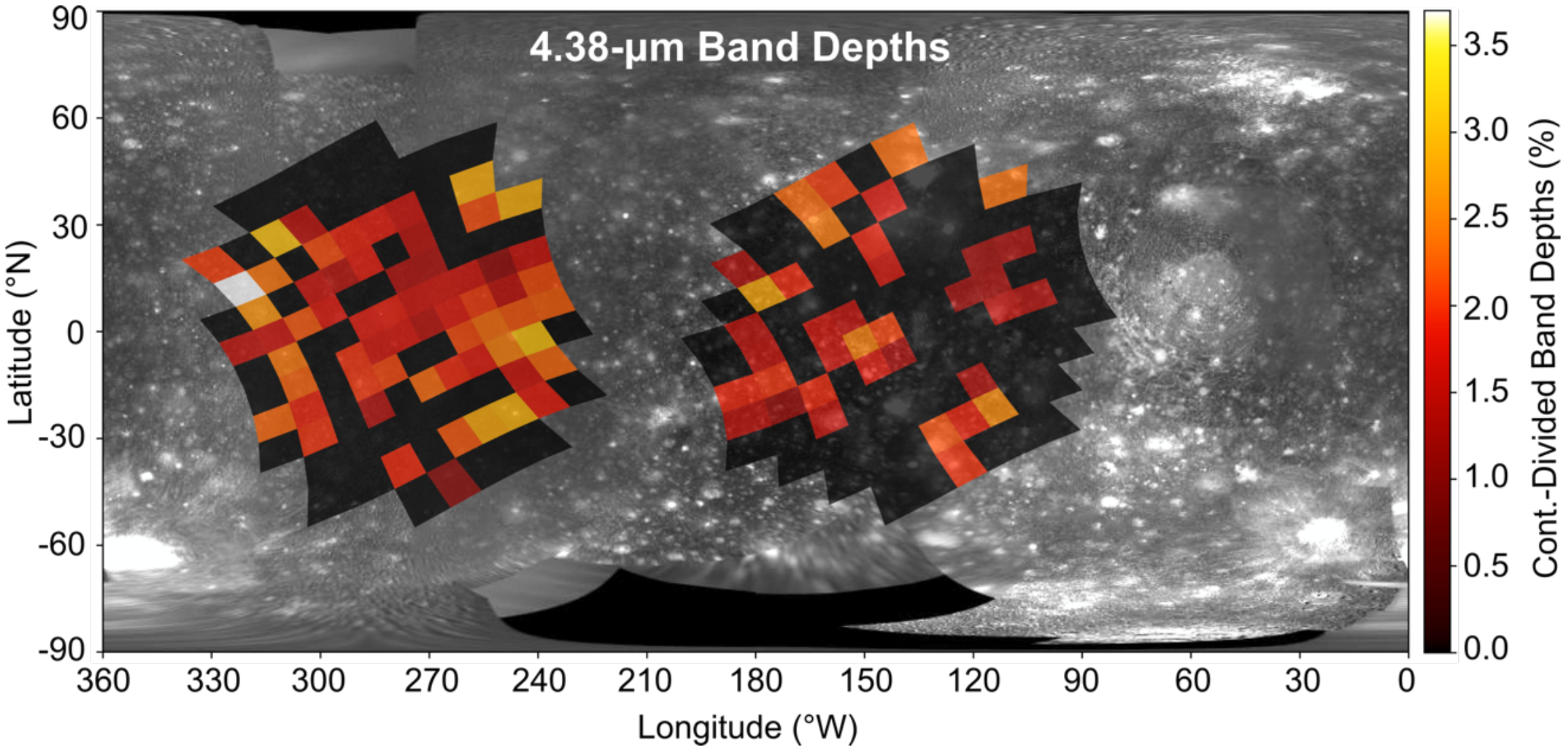

***Figure A4:*** *Band depth map for the 'peak' component of the 4.38-µm $^{13}CO_2$ band, highlighting the minor concentration of this component at low latitudes on Callisto's trailing hemisphere. The band center of this feature shows minimal variation, and it is very close to 4.38 µm in all spaxels where it was detected. Some spaxels do not show a peak component (black).*

*A.3 Description of possibly spurious and missing features*

In this appendix, we investigate whether Callisto's 4.38-μm band could be an artifact. We also describe several weak features that may result from artifacts or incompletely removed stellar lines. Finally, we detail absorption features that were detected in other datasets but are not observed in the integrated NIRSpec data of Callisto.

*Is Callisto's 4.38-μm band an artifact?* It has been suggested that a broad feature centered near 4.38 μm in NIRSpec G395H reflectance spectra of Ganymede could be a calibration artifact (Bockelée-Morvan et al., 2023). A similar 4.38-μm feature has been detected in NIRSpec data of Io, collected during Jupiter eclipse when Io's spectral properties should be dominated by thermal emission (de Pater et al., 2023). Saturn's A ring also exhibits a subtle 4.37-μm feature in NIRSpec PRISM mode data, which may be a pipeline calibration issue that contributes a wide and subtle 'dip' in high S/N data reduced using a solar reference spectrum (Hedman et al, 2023). In contrast, targets observed by NIRSpec that are flux calibrated using stars observed by NIRSpec (*i.e.*, Solar System object spectrum divided by G-type star spectrum) often do not exhibit the same subtle dip as it is divided out.

To investigate this possibility, we divided the Callisto G395H data by G395H spectra of P330E, a well-characterized spectrophotometric calibration star (G0, Vmag 13.028 ± 0.004, *e.g.*, Bohlin et al., 2015). The resulting disk-integrated spectrum exhibits a weak feature near 4.35 μm that could result from $^{13}CO_2$ (Figure A5). The P330E-divided data are noisier and two other absorption features not seen in the solar-model-divided version of the Callisto spectrum appear near 4.43 and 4.85 μm, making the validity of the 4.35-μm feature more difficult to assess.

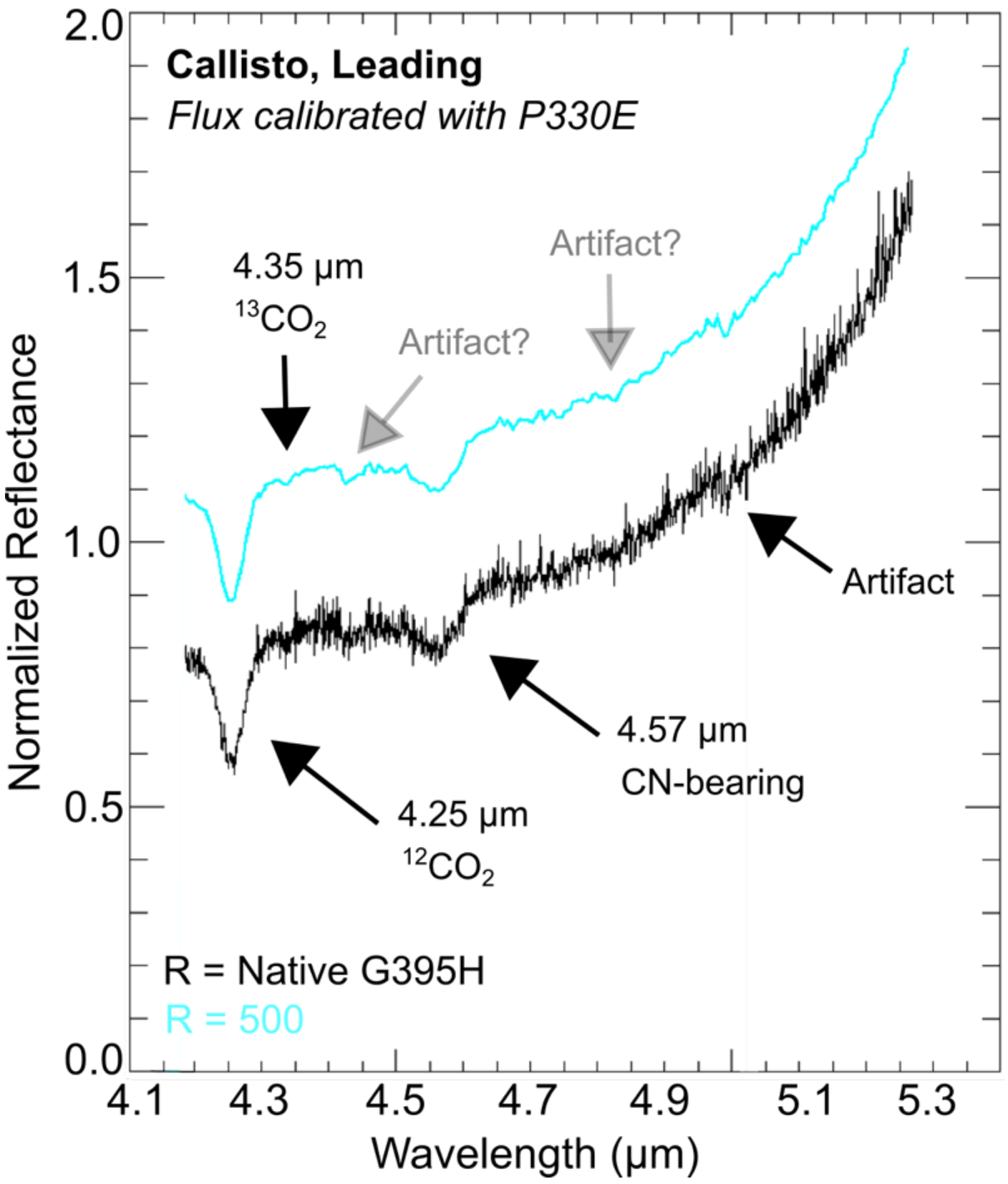

***Figure A5:*** *Native resolution integrated G395H (nrs2) spectrum of Callisto (R ~ 3000 at 4.3 μm, black) and a smoothed version of the same spectrum (R = 500, cyan), both of which were flux calibrated using P330E, with no thermal correction applied. Error bars have been suppressed for clarity. Features that are identified in both the P330E and solar model calibrated versions are bolded (see Figure 4 for solar model calibrated disk-integrated spectra). Features that are not detected in the solar-model-calibrated spectra are highlighted by question marks. The signal-to-noise is lower in this version likely because P330E (Vmag ~ 13) is considerably fainter than Callisto (Vmag ~ 6), making assessment of subtle features more challenging than in the higher quality versions shown in e.g., Figure 4.*

NIRSpec G395H data of Europa, which were reduced using a similar solar model to the Callisto data shown in *e.g.*, Figure 4, also exhibit a 4.38-µm band attributed to $^{13}CO_2$ (Villanueva et al., 2023a). Europa's 4.38-µm feature is considerably weaker than the broad feature detected on Ganymede and Callisto (non-Io-subtracted data), but it is comparable to the band strength of the 4.38-µm feature in Callisto's Io-subtracted data (Figure A6). Furthermore, G395H data of Enceladus, reduced using a similar solar model to the Galilean moons, do not exhibit $^{12}CO_2$ or $^{13}CO_2$ features (Villanaueva et al., 2023b), indicating that the 4.38-µm band likely requires the additional presence of a 4.25-µm $^{12}CO_2$ feature, consistent with absorption by $^{13}CO_2$. Of note, the S/N of the Enceladus data is lower than the Galilean moons and a weak ($\lesssim$ 2% band depth) calibration artifact could be present and obscured.

We think the most likely explanation is that Callisto's 4.38-µm band results from $^{13}CO_2$ molecules on its surface, as suggested to explain a 4.36-µm band identified in NIMS data (McCord et al., 1998a). It is possible that a calibration artifact is also contributing to Callisto's 4.38-µm band, artificially enhancing the base component of this feature, thereby increasing its band area (described in section 4.3). The Io-subtracted spectra still exhibit weak 4.38-µm absorption bands (Figure A6), consistent with the presence of subtle $^{13}CO_2$ features. The resulting $^{12}C/^{13}C$ ratios (~50) are within 2σ of the values calculated for Iapetus, the lower range of most comets, and other icy objects that exhibit terrestrial-like ratios.

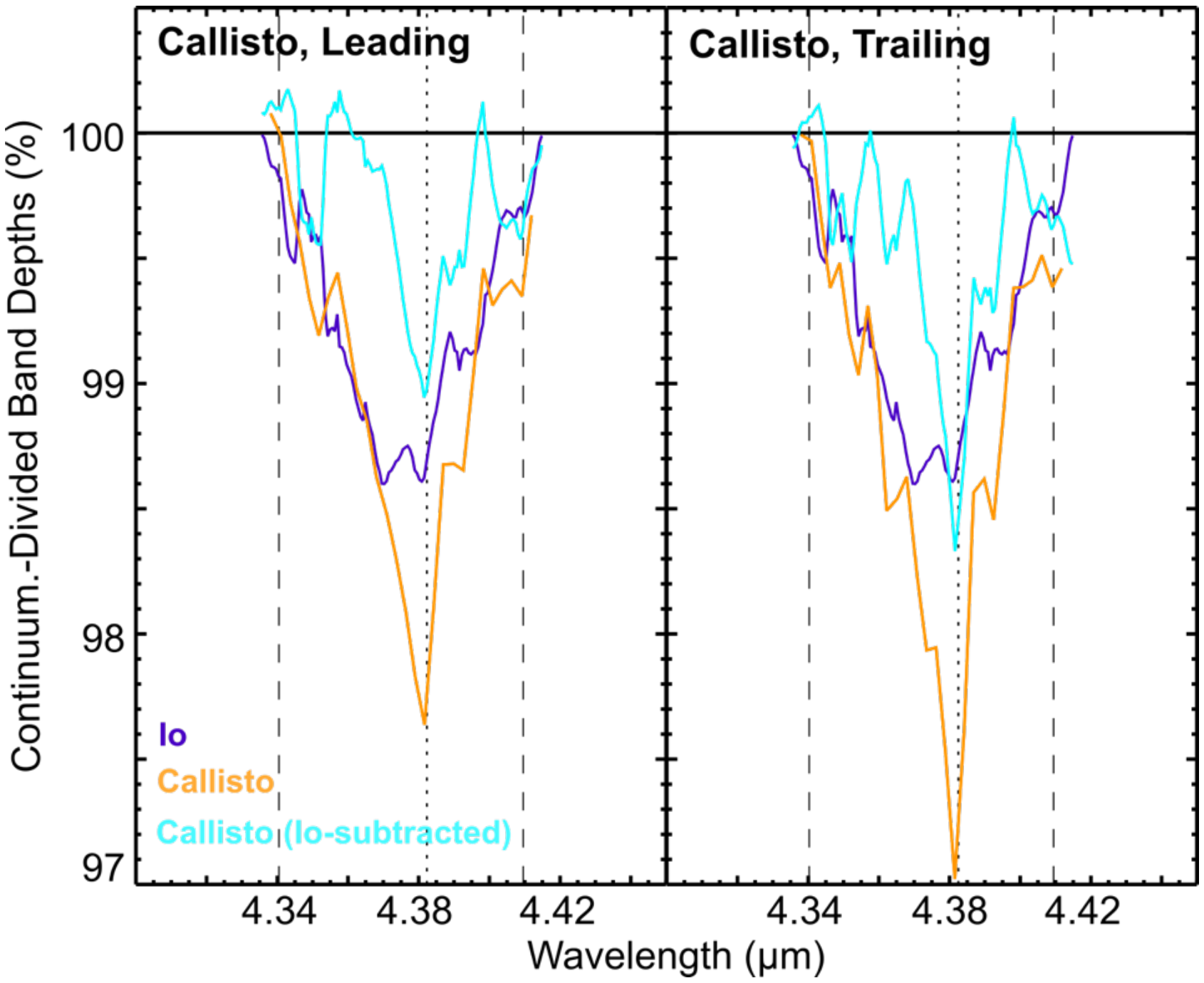

***Figure A6:*** *Continuum-divided, integrated G395H spectra of Io (purple) and non-Io-subtracted (orange) and Io-subtracted (cyan) integrated G395H spectra of Callisto. Each continuum-divided spectrum has been binned to an R of 500, and error bars have been suppressed for clarity.*

*Other possibly spurious features:* The NIRSpec data also show a narrow absorption feature centered near 4.30 µm that is stronger on Callisto's trailing side compared to its leading side (Table 1, Figure 4). This narrow band coincides with the wavelength position of the $^{16}O^{12}C^{18}O$ isotope of $CO_2$ measured in the laboratory that can form in response to irradiation of $H_2^{18}O$ mixed with carbonaceous material (*e.g.*, Mennella et al., 2004). However, there is a fairly strong solar line that is a very close match in wavelength position and band profile, and perhaps the weak 4.30-µm feature we have identified is a residual solar feature.

The integrated spectra exhibit several absorption features between 5 and 5.3 µm, that have not been previously identified, including a prominent feature near 5 µm. Given that the spectral structure beyond 5 µm exhibits a mostly unchanging morphology across Callisto's disk, its proximity to the long wavelength edge of the G395H detector, and its non-detection in older datasets, we suspect that these features are spurious. The same conclusion was reached for a very similar 5 µm feature that was detected in NIRSpec data of Ganymede, but not in data collected with JWST's Mid InfraRed Instrument (MIRI, 5 – 28 µm) (Bockelée-Morvan et al., 2023).

Nevertheless, we cannot completely rule out the possibility that some of the structure beyond 5 µm is real, and we briefly describe this possibility here. The prior non-detection of these features is unsurprising given that the sensitivity of NIMS is very low at wavelengths >5 µm and ground-based datasets are often swamped by atmospheric contamination. If these features are real, identification of possible species that might be contributing to them is difficult given that laboratory spectra spanning 5 to 5.3 µm typically do not exhibit many diagnostic features for C, O, and/or S-bearing species. One possibility is $C_2O$ that can exhibit an absorption feature near 5 µm in cryogenic irradiation experiments with C, O, and S bearing ices (*e.g.*, Ferrante et al., 2008). Additionally, higher-order hydrocarbons ($C_xH_x$) like paraffin can exhibit features in this wavelength range (Clark et al., 2009), which could hypothetically be contributing to dark material on Callisto. We leave deeper investigation of the putative spectral structure between 5 to 5.3 µm and the 4.30-µm band for future work.

*Detected in other datasets but absent from integrated NIRSpec data:* An absorption feature centered near 3.05 µm has been detected in some ground-based spectra of Callisto and attributed to $NH_4$-bearing compounds (Calvin and Clark, 1993). Additionally, some NIMS spectra show subtle features across the 2.8 to 3.1 µm region, which were attributed to O-H stretching modes in hydrated minerals (summarized in Figure 17.3 in Moore et al., 2004) and perhaps also contribute to the 3.05-µm feature detected in ground-based data. Although we do not detect a 3.05-µm feature in the integrated NIRSpec spectra, some individual spaxels show structure in this wavelength range that hint at the presence of another, non-$H_2O$ ice absorber.

Data collected by Galileo's Ultraviolet Spectrometer (UVS) suggest a minor amount of hydrogen peroxide ($H_2O_2$) may be present on Callisto (Hendrix et al., 1999), possibly manifesting as weak, broad features and reddish spectral slopes at wavelengths < 0.4 µm (Johnson and Quickenden, 1997). A prominent feature near 3.505 µm is attributed to $H_2O_2$ on Europa (*e.g.*, Carlson et al., 1999; Villanueva et al., 2023a) and Ganymede (Trumbo et al., 2023), but this feature was not detected on Callisto in NIMS or ground-based datasets. Thus, we think the subtle 3.51-µm band seen in NIRSpec data of Callisto is best matched by CH-bearing organics (see section 4.5), as were previously suggested to explain a 3.5-µm feature in NIMS data (McCord et al., 1998a).

NIMS detected a broad feature centered near 3.88 µm, which was attributed to carbonic acid ($H_2CO_3$) (*e.g.*, Johnson et al., 2004), as well as hydrogen sulfide ($H_2S$) (McCord et al., 1997, 1998a). The integrated NIRSpec spectra do not show convincing evidence for a 3.88-µm feature, albeit some individual spaxels show hints of a broad absorption band between 3.85 and 3.9 µm. These results are similar to the ambiguous detection of this feature in ground-based data, where only a narrow and weak 3.88-µm feature was noted, possibly more consistent with a residual telluric band or solar lines than a real feature (Cartwright et al., 2020). One possibility is that the 3.88-µm band detected by NIMS results from constituents that are spatially constrained to localized deposits that do not contribute meaningfully at the spatial scale of JWST or ground-based datasets. Additionally, laboratory experiments demonstrate that continued irradiation of $H_2CO_3$, after its formation from $H_2O$ ice mixed with $CO_2$, recycles $H_2CO_3$ back into its parent molecules, along with solid C (Strazzulla et al., 2023), which might limit the abundance of this molecule on Callisto and the other Galilean moons.

Another prominent band detected by NIMS is centered between 4.02 to 4.05 μm (McCord et al., 1997), which was confirmed by ground-based observations (Calvin and Clark, 1993; Cartwright et al., 2020). The 4.02-μm feature has been attributed to a range of species, including $SO_2$ (McCord et al., 1998a), carbonates (Johnson et al., 2004), hydrogen disulfide ($H_2S_2$) and/or disulfanide ($HS_2$), and S allotropes (Cartwright et al., 2020). The 4.02-μm band is absent from the integrated NIRSpec spectra because it overlaps the full range of the G395H's unrecoverable wavelength gap (4 – 4.2 μm). Nevertheless, the 4.02-μm band is observed in some individual spaxels covering Callisto's leading (18) and trailing (15) hemisphere, confirming that the feature is present. Comparison of these spaxels demonstrates that the 4.02-μm band is stronger on Callisto's leading hemisphere compared to its trailing hemisphere (Table 1, Figure 4), consistent with the hemispherical distribution measured with ground-based data (Cartwright et al., 2020). Analyses of individual NIRSpec spaxels that capture the 3.05-μm feature and the 4.02-μm band, and hint at the presence of the 3.88-μm feature, are beyond the scope of this project and left for future work.